\documentclass[twocolumn]{aastex62}

\usepackage{amsmath}
\usepackage{graphicx}   
\usepackage{bm}         
\usepackage{natbib}
\usepackage{etoolbox}
\usepackage{rotating}
\usepackage{array}
\usepackage{booktabs}
\usepackage{amssymb}
\usepackage{multirow}
\usepackage[symbol*]{footmisc}
\renewcommand{\thefootnote}{\fnsymbol{footnote}}
\usepackage{bold-extra}

\makeatletter
\patchcmd{\NAT@citex}
  {\@citea\NAT@hyper@{%
     \NAT@nmfmt{\NAT@nm}%
\hyper@natlinkbreak{\NAT@aysep\NAT@spacechar}{\@citeb\@extra@b@citeb}%
     \NAT@date}}
  {\@citea\NAT@nmfmt{\NAT@nm}%
   \NAT@aysep\NAT@spacechar\NAT@hyper@{\NAT@date}}{}{}

\patchcmd{\NAT@citex}
  {\@citea\NAT@hyper@{%
     \NAT@nmfmt{\NAT@nm}%
\hyper@natlinkbreak{\NAT@spacechar\NAT@@open\if*#1*\else#1\NAT@spacechar\fi}%
       {\@citeb\@extra@b@citeb}%
     \NAT@date}}
  {\@citea\NAT@nmfmt{\NAT@nm}%
\NAT@spacechar\NAT@@open\if*#1*\else#1\NAT@spacechar\fi\NAT@hyper@{\NAT@date}}
  {}{}
\makeatother

\makeatletter
\DeclareRobustCommand{\textsupsub}[2]{{%
  \m@th\ensuremath{%
    ^{\mbox{\fontsize\sf@size\z@#1}}%
    _{\mbox{\fontsize\sf@size\z@#2}}%
  }%
}}
\makeatother

\newcommand{\lsun}{\mbox{\,$L_\odot$}}
\newcommand{\msun}{\mbox{\,$M_\odot$}}

\newcommand{\kms}{\mbox{\,km\,s$^{-1}$}}

\newcommand{\lbol}{\mbox{$L_{\rm bol}$}}
\newcommand{\tbol}{\mbox{$T_{\rm bol}$}}

\newcommand{\methylformate}{\mbox{HCOOCH$_{3}$}}

\newcommand{\methanol}{\mbox{CH$_{3}$OH}}

\newcommand{\ethanol}{\mbox{C$_{2}$H$_{5}$OH}}
\newcommand{\acetaldehyde}{\mbox{CH$_{3}$CHO}}

\newcommand{\hcop}{\mbox{HCO$^{+}$}}
\newcommand{\water}{\mbox{H$_{2}$O}}

\newcommand{\source}{IRAS 15398$-$3359}

\shorttitle{}
\shortauthors{Yang et al.}

\begin{document}

\title{CORINOS I: JWST/MIRI Spectroscopy and Imaging of a Class 0 protostar \source}

\author[0000-0001-8227-2816]{Yao-Lun Yang}
\affiliation{RIKEN Cluster for Pioneering Research, Wako-shi, Saitama, 351-0198, Japan}
\affiliation{Department of Astronomy, University of Virginia, Charlottesville, VA 22904, USA}

\author[0000-0003-1665-5709]{Joel D. Green}
\affiliation{Space Telescope Science Institute, Baltimore, 3700 San Martin Dr., MD 21218, USA}

\author[0000-0001-7552-1562]{Klaus M. Pontoppidan}
\affiliation{Space Telescope Science Institute, Baltimore, 3700 San Martin Dr., MD 21218, USA}


\author{Jennifer B. Bergner}
\affiliation{University of Chicago Department of the Geophysical Sciences, Chicago, IL 60637, USA}
\altaffiliation{NASA Sagan Fellow}

\author[0000-0003-2076-8001]{L. Ilsedore Cleeves}
\affiliation{Department of Astronomy, University of Virginia, Charlottesville, VA 22904, USA}

\author[0000-0001-5175-1777]{Neal J. Evans II}
\affiliation{Department of Astronomy, The University of Texas at Austin, Austin, TX 78712, USA}

\author[0000-0001-7723-8955]{Robin T. Garrod}
\affiliation{Departments of Chemistry and Astronomy, University of Virginia, Charlottesville, VA, 22904, USA}

\author[0000-0002-4801-436X]{Mihwa Jin}
\affiliation{Astrochemistry Laboratory, Code 691, NASA Goddard Space Flight Center, Greenbelt, MD 20771}
\affiliation{Department of Physics, Catholic University of America, Washington, DC 20064, USA}

\author{Chul Hwan Kim}
\affiliation{Department of Physics and Astronomy, Seoul National University, 1 Gwanak-ro, Gwanak-gu, Seoul 08826, Korea}

\author{Jaeyeong Kim}
\affiliation{Korea Astronomy and Space Science Institute, 776 Daedeok-daero, Yuseong-gu Daejeon 34055, Republic of Korea}

\author{Jeong-Eun Lee}
\affiliation{Department of Physics and Astronomy, Seoul National University, 1 Gwanak-ro, Gwanak-gu, Seoul 08826, Korea}

\author[0000-0002-3297-4497]{Nami Sakai}
\affiliation{RIKEN Cluster for Pioneering Research, Wako-shi, Saitama, 351-0198, Japan}

\author[0000-0002-5171-7568]{Christopher N. Shingledecker}
\affiliation{Department of Physics and Astronomy, Benedictine College, Atchison, KS, 66002, USA}

\author{Brielle Shope}
\affiliation{Department of Chemistry, University of Virginia, 409 McCormick Rd, Charlottesville, VA, 22904, USA}

\author[0000-0002-6195-0152]{John J. Tobin}
\affiliation{National Radio Astronomy Observatory, 520 Edgemont Rd., Charlottesville, VA 22903, USA}

\author[0000-0001-7591-1907]{Ewine F. van Dishoeck}
\affiliation{Leiden Observatory, Leiden University, Netherlands}
\affiliation{Max Planck Institute for Extraterrestrial Physics, Garching, Germany}

\correspondingauthor{Yao-Lun Yang}
\email{yaolunyang.astro@gmail.com}

\begin{abstract}
  The origin of complex organic molecules (COMs) in young Class 0 protostars has been one of the major questions in astrochemistry and star formation.  While COMs are thought to form on icy dust grains via gas-grain chemistry, observational constraints on their formation pathways have been limited to gas-phase detection. Sensitive mid-infrared spectroscopy with JWST enables unprecedented investigation of COM formation by measuring their ice absorption features. 
  Mid-infrared emission from disks and outflows provide complementary constraints on the protostellar systems.
  We present an overview of JWST/MIRI MRS spectroscopy and imaging of a young Class 0 protostar, \source, and identify several major solid-state absorption features in the 4.9--28 \micron\ wavelength range.  These can be attributed to common ice species, such as \water, \methanol, NH$_3$, and CH$_4$, and may have contributions from more complex organic species, such as \ethanol\ and CH$_3$CHO.  In addition to ice features, the MRS spectra show many weaker emission lines at 6--8 \micron, which are due to warm CO gas and water vapor, possibly from a young embedded disk previously unseen. Finally, we detect emission lines from [Fe\,\textsc{ii}], [Ne\,\textsc{ii}], [S\,\textsc{i}], and H$_2$, tracing a bipolar jet and outflow cavities. MIRI imaging serendipitously covers the south-western (blue-shifted) outflow lobe of \source, showing four shell-like structures similar to the outflows traced by molecular emission at sub-mm wavelengths. This overview analysis highlights the vast potential of JWST/MIRI observations and previews scientific discoveries in the coming years.
\end{abstract}

\section{Introduction}

In recent years, complex organic molecules (COMs), first detected in high-mass cores \citep{1985ApJS...58..341S,1986ApJS...60..357B,1987ApJ...315..621B}, have been routinely detected in the gas-phase in low-mass protostellar cores, suggesting extensive chemical evolution at the early stage of low-mass star formation \citep[e.g.,][]{1995ApJ...447..760V,ceccarelli2007extreme,2020ARAA..58..727J,2022arXiv220613270C}. 
These low-mass cores are often called ``hot corinos'' \citep{2003ApJ...593L..51C,2004ASPC..323..195C,2004ApJ...615..354B}.
The COMs, commonly defined as organic molecules with six or more atoms \citep{2009ARAA..47..427H}, could be the precursors of pre-biotic molecules \citep[e.g.,][]{2020AsBio..20.1048J}. Solar system objects, such as comets, also show abundant COMs \citep{2019ARAA..57..113A}; and in some cases, the COM abundances match those measured in protostellar cores, hinting at a chemical connection from protostars to planetary systems \citep{2000AA...353.1101B,2019MNRAS.490...50D,2019ESC.....3.2659B} Thus, the origin of the rich organic chemistry in the protostellar stage is of great interest in characterizing the chemical environment of planet-forming disks.

Current models predict that a combination of gas-phase and ice-phase processes (i.e., `gas-grain chemistry') is responsible for COM formation in protostellar environments \citep[e.g.,][]{2008ApJ...682..283G,2014ApJ...791....1T,2018ApJ...869..165L,2018MNRAS.474.2796Q,2018ApJ...854..116S,2020ApJ...897..110A}.  These models generally require a warm-up phase during which the elevated temperature enables efficient reactions via diffusion.  In addition to the formation of COMs in the ice phase, gas-phase reactions following sublimation of simpler ice molecules may contribute to the production of several COMs \citep{2015MNRAS.449L..16B,2018ApJ...854..135S,2020MNRAS.499.5547V}.  Laboratory experiments show that COMs can also be formed on icy surfaces even at low temperature \citep{2017ApJ...842...52F,2016MNRAS.455.1702C,2019ApJ...874..115B,2019ESC.....3..986Q}.  Extended distributions of COMs in cold prestellar cores further suggest ongoing formation of COMs in the ice-phase \citep{2016ApJ...830L...6J,2017ApJ...842...33V,2020ApJ...891...73S,2022ApJ...927..213P}. To reconcile the presence of COMs at low temperature, a modified gas-grain chemical model that includes non-diffusive reactions at low temperature has been proposed \citep{2020ApJS..249...26J,2022ApJS..259....1G}.

Recent surveys show that gas-phase COM emission is common, but not ubiquitous, around Class 0/I protostars, with detection fractions around half \citep{2019MNRAS.483.1850B,2019ESC.....3.1564B,2020AA...635A.198B,2020AA...639A..87V,2021ApJ...910...20Y,2021AA...650A.150N,2022ApJ...929...10B,2022ApJ...927..218H}.  It remains unknown why some sources show rich emission of gas-phase organics and others do not.  It may be a true chemical effect, with some sources having low ice-phase COM reservoirs due to their environmental/evolutionary conditions.  Another possibility is that COMs are only efficiently sublimated into the gas phase in a subset of sources.
Disk shadowing can effectively lower the temperature in the envelope, leading to inefficient desorption and thus low abundance of gaseous COMs, hence non-detection \citep{2022AA...663A..58N}.
Moreover, high dust optical depth could suppress the COM emission at sub-mm wavelengths \citep{2020ApJ...896L...3D,2022AA...663A..58N}.
Disentangling these scenarios requires an understanding of COM abundances in the ice phase.  Therefore, mid-infrared spectroscopy of organic ice features offers an avenue to understand the origin and nature of complex molecule formation in protostars.

Outflows are ubiquitously associated with protostellar cores.  The clearance of an outflow cavity and the accretion activity that is tightly related to outflows regulates the thermal structure of the envelope as well as the photochemistry along the cavity wall, thus affecting the abundance of COMs in both gas- and ice-phase \citep[e.g.,][]{2012AA...537A..55V,2014MNRAS.445..913D,2015MNRAS.451.3836D}.  At mid-infrared wavelengths, rotationally excited H$_2$ lines and ionic forbidden lines trace the shocked gas and jets in outflow cavities \citep[e.g.,][]{2010AA...519A...3L}.  Furthermore, ro-vibrational CO lines and water vapor emission at $\sim$4--6 \micron\ highlight the shocked gas at the base of outflows and/or at the disk surface, constraining the physical conditions of outflows and disks \citep[e.g.,][]{2011AA...533A.112H,2022AJ....164..136S}.

The CORINOS (COMs ORigin Investigated by the Next-generation Observatory in Space) program measures the ice composition of four isolated Class 0 protostars with JWST (program 2151, PI: Y.-L. Yang). The program aims to determine the abundances of ice species with radiative transfer and chemical modeling to constrain the formation and evolution of COMs. The full sample consists of two protostars whose gas-phase spectra are known to exhibit rich COM features, B335 and L483, and two protostars with little emission of gas-phase COMs, \source\ and Ser-emb 7 \citep{2009ApJ...697..769S,2016ApJ...830L..37I,2017ApJ...837..174O,2019ESC.....3.1564B,2019AA...629A..29J}.  Each pair represents low- ($\sim1$ \lsun) and high-luminosity ($\sim10$ \lsun) protostars. This work presents initial results from the first observation of \source.

In this paper, we present JWST/MIRI observations of \source, highlighting several new mid-infrared ice features, likely associated with COMs, as well as emission lines and outflows detected in both spectroscopy and imaging. In Section\,\ref{sec:observations}, we describe our JWST/MIRI observing program and data reduction. In Section\,\ref{sec:ice}, we show the extracted 1D MRS spectra and identify absorption features in the spectra along with possible contributing ice species.  Section\,\ref{sec:water_vapor} presents the detection of warm water vapor and CO emission, which may originate in a young protoplanetary disk.  Section\,\ref{sec:outflows} shows the south-western outflow of \source\ in MIRI imaging and presents detected emission lines, most of which trace the outflows and jets.  Lastly, in Section\,\ref{sec:conclusions}, we highlight the findings with this first analysis of JWST/MIRI spectra of a Class 0 protostar.

\subsection{\source}
\source\ (also known as B228) is a Class 0 protostar located in the Lupus I Molecular Cloud \citep{1989PASP..101..816H,2007ApJ...667..288C} at a distance of 154.9$^{+3.2}_{-3.4}$ pc \citep{2020AA...643A.148G}. It has a bolometric luminosity (\lbol) of 1.5 \lsun\ and a bolometric temperature (\tbol) of 68$\pm$27 K\citep{2018ApJ...860..174Y,2021AA...648A..41V}.
\source\ has drawn astrochemical interest because of its abundant warm carbon-chain molecules (CCMs), which suggests an active Warm Carbon-Chain Chemistry \citep[WCCC;][]{2009ApJ...697..769S} and chemical signatures of episodic accretion \citep[e.g.,][]{2013ApJ...779L..22J}. In the WCCC scenario, abundant CH$_4$ ice, which may form in the prestellar stage, is sublimated as the temperature increases due to accretion heating, leading to an elevated abundance of carbon carriers available for the formation of CCMs \citep{2008ApJ...672..371S,2008ApJ...674..984A}. High UV illumination at the prestellar stage, may explain abundant carbon-chain molecules in protostars \citep{2016AA...592L..11S}. On the other hand, only a few emission lines of complex organic molecules (COMs) have been detected despite its rich CCMs (Okoda et al. in prep.). The location of the envelope water snowline inferred from \hcop\ , as well as by detection of HDO, is larger than the current luminosity of \source\ \citep{2013ApJ...779L..22J,2016AA...595A..39B}, suggesting a higher luminosity in the last 100--1000 years, perhaps due to an accretion burst.   Moreover, the ice features of \source\ were studied in the Spitzer ``c2d'' (Cores to Disks) survey, where common species, such as \water, CO$_2$, CH$_4$, and \methanol, were identified \citep{2008ApJ...678..985B,2008ApJ...678.1005P,2008ApJ...678.1032O,2010ApJ...718.1100B}.

\source\ is associated with a compact disk, although poorly constrained by observations. \citet{2017ApJ...834..178Y} estimated a centrifugal radius ($R_\text{c} = \frac{j^2}{GM_\star}$, where $j$ is the specific angular momentum) of 20$^{+50}_{-20}$ au by fitting the C$^{18}$O emission.  With a similar method, \citet{2018ApJ...864L..25O} found the centrifugal barrier ($R_\text{cb} = \frac{j^2}{2GM_\star}$) at 40 au can explain the kinematics of the SO emission, which corresponds to a centrifugal radius of 80 au.  The estimated disk radii from both studies are consistent with considerable uncertainty due to the unresolved Keplerian rotation.  They also estimated a very low protostellar mass of only $\leq$0.01$^{+0.02}_{-0}$ \msun\ and 0.007$^{+0.004}_{-0.003}$ \msun\ by \citet{2017ApJ...834..178Y} and \citet{2018ApJ...864L..25O}, respectively.  

The bipolar outflow of \source\ has a young dynamical age of $\sim 1000$\,yr, as measured from the CO outflow \citep{2015AA...576A.109Y,2016AA...587A.145B}.  The outflow consists of a wide-angle wind-driven outflow and jet-driven bow-shocks \citep{2016AA...587A.145B,2017ApJ...834..178Y}.  \citet{2020ApJ...900...40O} show compact emission of H$_2$CO in the outflow identified with a Principal Component Analysis, suggesting a shock-induced origin. \citet{2021AA...648A..41V} further showed evidence of a precessing episodic jet-driven outflow with four ejections separated by 50--80 years. Recently, \citet{2021ApJ...910...11O} found an arc-like structure perpendicular to the known outflow, which they interpreted as shocked gas due to a previously launched secondary outflow.

\section{Observations}
\label{sec:observations}
The protostar \source\ was observed with the Mid-InfraRed Instrument \citep[MIRI;][]{2015PASP..127..584R,2015PASP..127..595W} onboard JWST on 2022 July 20, as part of program 2151 (PI: Y.-L. Yang). The observations used the Medium Resolution Spectroscopy (MRS) mode, which is equipped with four Integral Field Units (IFU) that observed the target simultaneously using dichroics. These IFUs are often referred as ``channels'', where channels 1, 2, 3, and 4 cover 4.9--7.65, 7.51--11.71, 11.55--18.02, and 17.71--28.1 \micron, respectively.  Each channel is covered by the same three grating settings, which are also called ``sub-bands''.  Thus, an exposure with only one grating setting results in four discontinuous spectra.  A full 4.9--28 \micron\ coverage requires observations with three grating settings, resulting in twelve spectral segments.  The spectroscopic data were taken in SLOWR1 readout mode with a standard 4-point dither pattern. 

\source\ was observed with a pointing center on ($15^{\mathrm{h}}43^{\mathrm{m}}02.24^{\mathrm{s}}$, $-34^\circ{09}^\prime{06.7}^{\prime\prime}$) based on the sub-mm continuum peak from \citet{2014ApJ...795..152O} along with a dedicated background pointing centered on ($15^{\mathrm{h}}43^{\mathrm{m}}07.9^{\mathrm{s}}$, $-34^\circ{09}^\prime{01}^{\prime\prime}$). Recent Atacama Large Millimeter/submillimeter Array (ALMA) observations suggest a sub-mm continuum peak at ($15^{\mathrm{h}}43^{\mathrm{m}}02.2307^{\mathrm{s}}$, $-34^\circ{09}^\prime{06.99}^{\prime\prime}$) using the ALMA Band 6 observations taken on 2022 May 16 (2021.1.00357.S; PI: S. Notsu). The integration time is 1433.4 seconds for the SHORT(A) and LONG(C) sub-bands and 3631.3 seconds for the MEDIUM(B) sub-band. The MEDIUM(B) sub-band covers the 8.67--10.15 \micron\ range where the intensity is the lowest due to strong absorption of silicates.  Thus, we intentionally integrated longer with the MEDIUM(B) setting to achieve a sufficient signal-to-noise ratio (S/N) to characterize the ice features around the silicate feature.

The data were processed from the Stage 1 data files (\texttt{uncal}) using v1.7.2 of the JWST pipeline and CRDS context (\texttt{jwst\_0977.pmap}) from \texttt{https://jwst-crds-pub.stsci.edu/}. The dedicated background exposures were subtracted on the exposure level during Stage 2 of the pipeline.  The Stage 3 process includes \texttt{OutlierDetectionStep}, \texttt{ResidualFringeStep}, and \texttt{CubeBuildStep}.  The \texttt{ResidualFringeStep} task is included to correct for residual fringes that are not fully corrected by the application of a fringe flat, particularly in extracted point source spectra.  The fringe is suppressed in most sub-bands except for noticeable residuals in \texttt{ch3-long} around 10--12 \micron. The wavelength calibration is generally accurate to within $\sim$1 spectral resolution element \citep[$\sim$100 \kms;][]{2022arXiv220705632R}.

The protostar appears point-like in the MRS spectral cube. Thus, we extracted a 1D spectrum with an aperture ($R_{\rm ap}$) defined by the diffraction-limited beam size ($1.22\lambda/D$) so that the aperture increases with wavelength.  The aperture centers at the ALMA continuum peak ($15^{\mathrm{h}}43^{\mathrm{m}}02.2307^{\mathrm{s}}$, $-34^\circ{09}^\prime{06.99}^{\prime\prime}$).  We tested the spectral extraction with additional local background subtraction derived from an annulus outside the aperture; however, the resulting spectra appear to have more noise possibly because the extended outflow cavity complicates the determination of the true background.  Thus, we performed no additional background subtraction on the reduced spectral cubes.  Despite its point-like appearance, the source emission extends beyond the size of the diffraction-limited beam.  A 1D spectrum extracted with a small aperture results in inconsistent flux between several sub-bands due to the flux extended beyond the aperture.  Appendix\,\ref{sec:extraction} shows a detailed analysis of the extracted spectra with different apertures.  We find that a four-beam aperture provides a good balance between the flux agreement between sub-bands and noise.  All spectra show in this study are extracted with a four-beam aperture, $4\times1.22\lambda/D$, unless otherwise specified. We further matched the flux between channels by the ratio of median fluxes in the overlapping wavelengths by applying scale factors of order $\lesssim$16\%, starting from the shortest wavelength.  The scaled spectrum differs from the original spectrum by at most 16\%.

To estimate the RMS in the extracted 1D spectrum, we subtracted a Gaussian-smoothed baseline and calculated the RMS in the residual with respect to the smoothed baseline, which has a median of 0.8\%\ with a 1$\sigma$ range from 0.4--1.3\%.  The Gaussian width is chosen as 20 wavelength channels to approximate the baseline without noise and avoid smoothing out broad absorption features.  The RMS may be underestimated between 10 and 12 \micron, where the fringe residuals are not fully suppressed.

Simultaneous MIRI imaging was enabled along with the primary spectroscopic observations for astrometric registration. The simultaneous field is pointed off the MRS target, but the background observation happened to be arranged such that it covered the south-western outflow lobe of \source. The imaging fields were observed with FASTR1 readout pattern, in the F560W, F770W, and F1000W filters, with filter widths of 1.2, 2.2, and 2.0 \micron, respectively. The point spread function (PSF) full width at half maximum (FWHM) in these bands was measured to 0\farcs22, 0\farcs25, and 0\farcs32, respectively. The total exposure time was 1433.4, 1433.4, and 3631.3 seconds, the same as their spectroscopic counterparts. The Stage 3 products were generated by the standard pipeline obtained from the Barbara A. Mikulski Archive for Space Telescopes (MAST); the data were calibrated with \texttt{jwst\_0932.pmap} from \texttt{https://jwst-crds.stsci.edu/} without further re-processing. The RMS noise estimated from the standard deviation in an empty sky region is 2.3, 7.6, and 18.4 MJy sr$^{-1}$, respectively.

\section{Ice bands in the point source spectrum}
\label{sec:ice}
\begin{figure*}[htbp!]
  \centering
  \includegraphics[width=\textwidth]{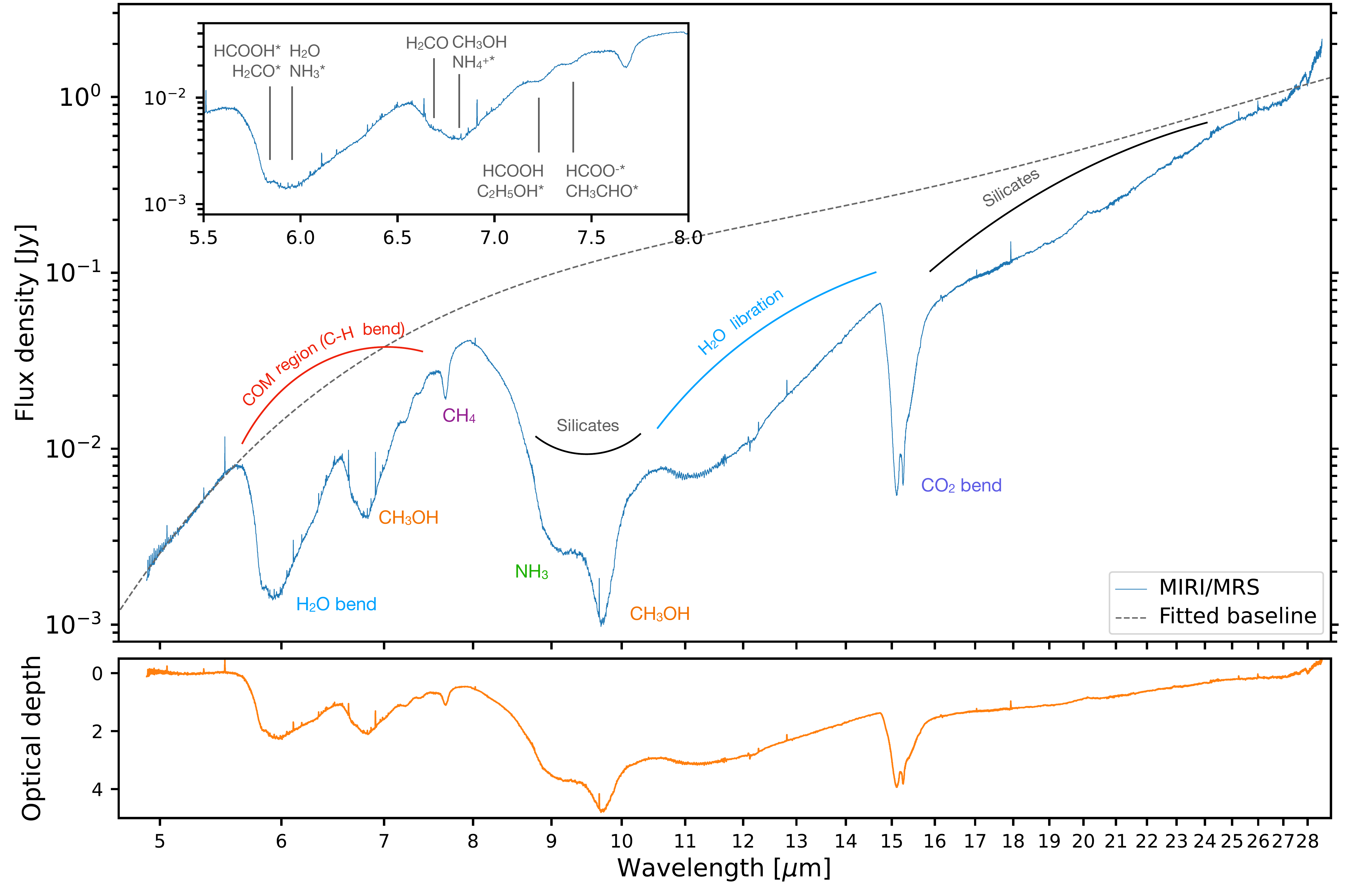}
  \caption{\textbf{Top}: Extracted MIRI MRS spectrum of the \source\ point source, with major solid-state features indicated. The wavelength axis is in logarithmic scale. The dashed line illustrates the fitted continuum. \textbf{Top (inset)}: Detail of the 5.5--8 \micron\ region from same spectrum  with secure and possible identifications labeled (see Table \ref{tbl:abs_features}). \textbf{Bottom}: The optical depth spectrum derived using the continuum shown in the top panel.}
  \label{fig:1d_spec}
\end{figure*}

The extracted MIRI MRS spectrum shows strongly increasing flux density with wavelength along with several absorption features, which is typical for embedded protostars (Figure\,\ref{fig:1d_spec}, top). All of the identified absorption features are due to ices and silicates. We estimate the large-scale continuum by fitting a fourth-order polynomial using the 5.05--5.15, 5.3--5.4, and 5.52--5.62 \micron\ range of the MRS spectrum and the 35--38 \micron\ range of the scaled Spitzer/IRS spectrum (see Figure\,\ref{fig:irs_comp}, right). Ideally the spectrum at the longest wavelengths, which is less affected by silicate and \water\ absorption, would be included for the continuum fitting.  However, the long wavelength end ($> 27.5$ \micron) of the MRS spectrum has higher noise and a steeper slope compared to the spectrum at 16--27 \micron; thus, we consider the $>27.5$ \micron\ spectrum as less reliably calibrated compared to the rest of the spectrum due to the rapid drop in MRS sensitivity at its longest wavelengths.  Including the Spitzer/IRS spectrum allows us to perform the continuum fitting at longer wavelengths ($> 30$ \micron).  The fitted continuum is consistent with the long wavelength end of the MIRI MRS spectrum.  Nonetheless, this fit has substantial systematic uncertainty depending on various factors, such as the choice of assumed absorption-free ranges and the functional form of the continuum. 
The qualitative analysis presented here serves to identify potential carriers of the ice bands, rather than to derive precise ice abundances.

Figure \ref{fig:1d_spec} (bottom) shows the optical depth spectrum, derived as $\tau= -{\rm ln}(F/C)$, where $F$ is the flux density and $C$ is the fitted continuum. We clearly detect the silicate band centered at 10\,$\mu$m, as well as the bending and libration modes of \water\ ice at 6 and 11--13\,$\mu$m, respectively. We also securely detect \methanol\ via the strong band at 9.7\,$\mu$m, supported by substructure at 6.8\,$\mu$m, CH$_4$ at 7.7$\mu$m, and CO$_2$ via its bending mode at 15.2\,$\mu$m. In addition, we highlight notable absorption features due to minor species that still have ambiguous identifications. The features and qualitative description of their shape are listed in Table\,\ref{tbl:abs_features}, where tentative identifications are marked with asterisks. In the following paragraphs, we discuss individual features. 

\subsection{Individual Features}
\label{sec:features}

\begin{deluxetable}{ccc}
\tablecaption{Notable ice features}
\label{tbl:abs_features}
\tablehead{
\colhead{Wavelength} & \colhead{Type} & \colhead{Identification} \\
($\mu$m) & & 
} 
\startdata
5.83        & single    & HCOOH*, H$_2$CO* \\
6           & multiple  & H$_2$O, NH$_3$* \\
6.7         & single    & H$_2$CO \\
6.8         & multiple  & \methanol, NH$_4^+$* \\
7.24        & single    & HCOOH, \ethanol* \\
7.41        & single    & HCOO$^-$*, CH$_3$CHO* \\
7.7         & single    & CH$_4$, SO$_2$*, \ethanol* \\
9           & single    & NH$_3$, \methanol*, \ethanol* \\
9.7         & single    & \methanol\ \\
11          & single/broad  & \water, \ethanol*, \\
            &           &\acetaldehyde*, \methylformate* \\
15.2        & multiple  & CO$_2$ \\
\enddata
\tablenotetext{*}{Potential/ambiguous identification}
\end{deluxetable}

\subsubsection{5.83 \micron\ feature: HCOOH*\footnote[0]{*Potential/ambiguous identification} and H$_2$CO*}
\label{sec:5.83}

This feature is likely due to the C=O stretching mode of HCOOH \citep{marechal1987ir,2007AA...470..749B} and/or H$_2$CO \citep{1993Icar..104..118S}. The feature is seen in the MIRI spectrum as a blue shoulder on the broad ($\sim$0.5 \micron) feature of the \water\ bending mode in the 5.8--6.3 \micron\ region \citep{1996AA...315L.333S}. 
\citet{2008ApJ...678..985B} measured the abundance of HCOOH as 1.9\%\ relative to \water\ using the 7.25 \micron\ feature of HCOOH, which we also detect (Section\,\ref{sec:7.24}).
Even if the identification of HCOOH is independently confirmed, both species could contribute to this C=O stretching mode at 5.8 \micron. In fact, \citet{2008ApJ...678..985B} showed that H$_2$CO can contribute no more than 10\%--35\%\ of this feature based on the non-detection of its absorption features at 3.34, 3.47, and 3.54 \micron\ in L-band spectra of other sources.  

\subsubsection{6 \micron\ feature: \water\ and NH$_3$*}
\label{sec:6}
The \water\ bending mode peaks at 6 \micron, dominating this feature \citep[e.g.,][]{2001AA...376..254K}. The N--H deformation mode of NH$_3$ at 6.16 \micron, whose umbrella mode at 9 \micron\ is detected (Section\,\ref{sec:9}), also contributes to this broad feature \citep{2008ApJ...678..985B}.
While the 6 \micron\ feature is detected in all low-mass protostars, the absorption from \water\ and NH$_3$ often underestimates the depth of this feature, suggesting additional contributions from unidentified species.

\subsubsection{6.7 \micron\ feature: H$_2$CO}
\label{sec:6.7}
\begin{figure*}[htbp!]
  \centering
  \includegraphics[width=\textwidth]{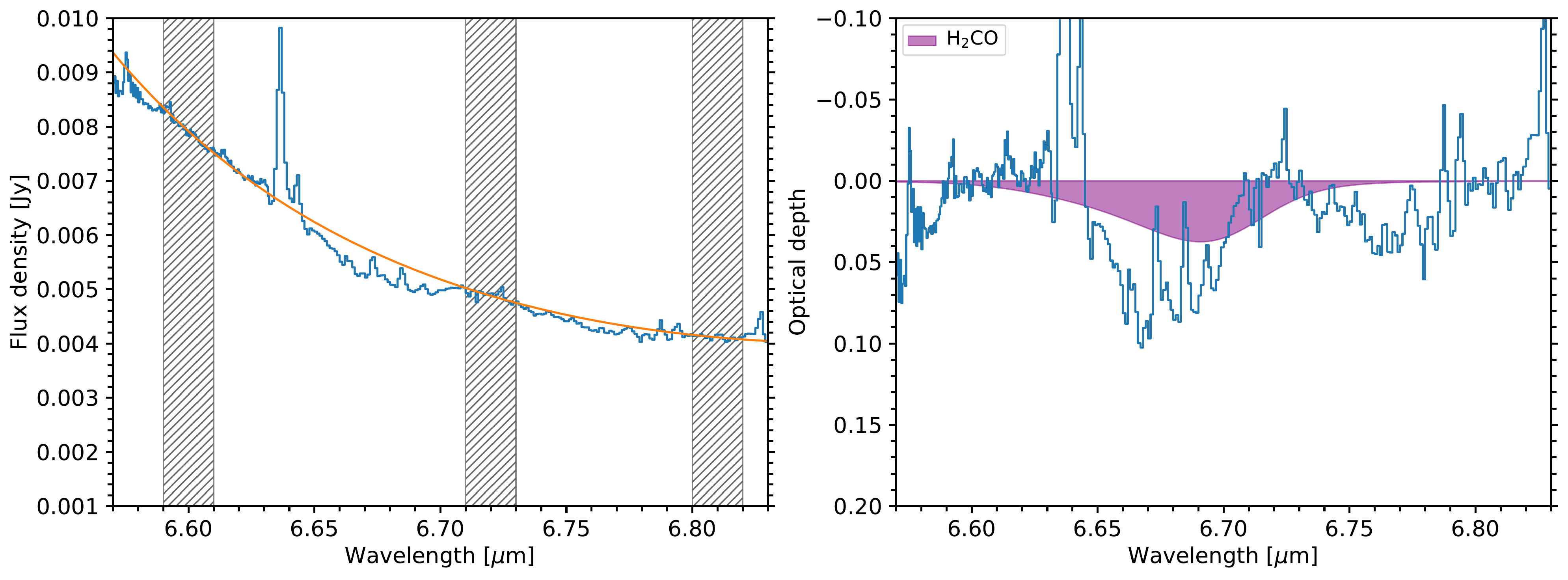}
  \caption{\textbf{Left}: The MIRI MRS spectrum of \source\ along with a second-order polynomial baseline fitted at 6.59--6.61 and 6.79--6.81 \micron\ (gray hatched regions). \textbf{Right}: The optical depth spectra of the shallow absorption at 6.68--6.72 \micron. The purple shaded region shows the scaled laboratory absorbance spectrum from \citet{1996AA...312..289G}. The narrow emission features are predominantly due to warm gas-phase water (Section\,\ref{sec:water_vapor}).}
  \label{fig:6-7}
\end{figure*}

We detect a shallow inflection on the blue side of the 6.8\,$\mu$m band (Section\,\ref{sec:6.8}). \citet{1993Icar..104..118S} reported that the C--H bending mode of H$_2$CO occurs at 6.68 \micron. In the c2d survey, \citet{2008ApJ...678..985B} put an upper limit of 15\%\ contribution from this bending mode to the absorption feature centered on 6.85 \micron. In our MIRI spectrum, the optical depth of this feature is $\sim0.05$ with a local baseline fitting (Figure\,\ref{fig:6-7}) and the overall optical depth of the entire 6.8\,$\mu$m band is $\sim1.5$, consistent with the suggested upper limit.

\subsubsection{6.8 \micron\ feature: \methanol\ and NH$_4^+$*} 
\label{sec:6.8}
This feature is ubiquitous in icy sightlines toward protostars and in the dense interstellar medium, and \source\ is no exception. Its position and shape is broadly consistent with the C--H bending mode of \methanol\ \citep{2008ApJ...678..985B}. \citet{2003AA...398.1049S} proposed that NH$_4^+$ could be a significant contributor; however, the identification of NH$_4^+$, based on the 6.8\,$\mu$m band alone remains debated, while \methanol\ can be confirmed given the observation of the corresponding C--O stretching mode at 9.75 \micron\ in \source\ (Section\,\ref{sec:9.7}).

\subsubsection{7.24 \micron\ feature: HCOOH and \ethanol*} 
\label{sec:7.24}
\begin{figure*}[htbp!]
  \centering
  \includegraphics[width=\textwidth]{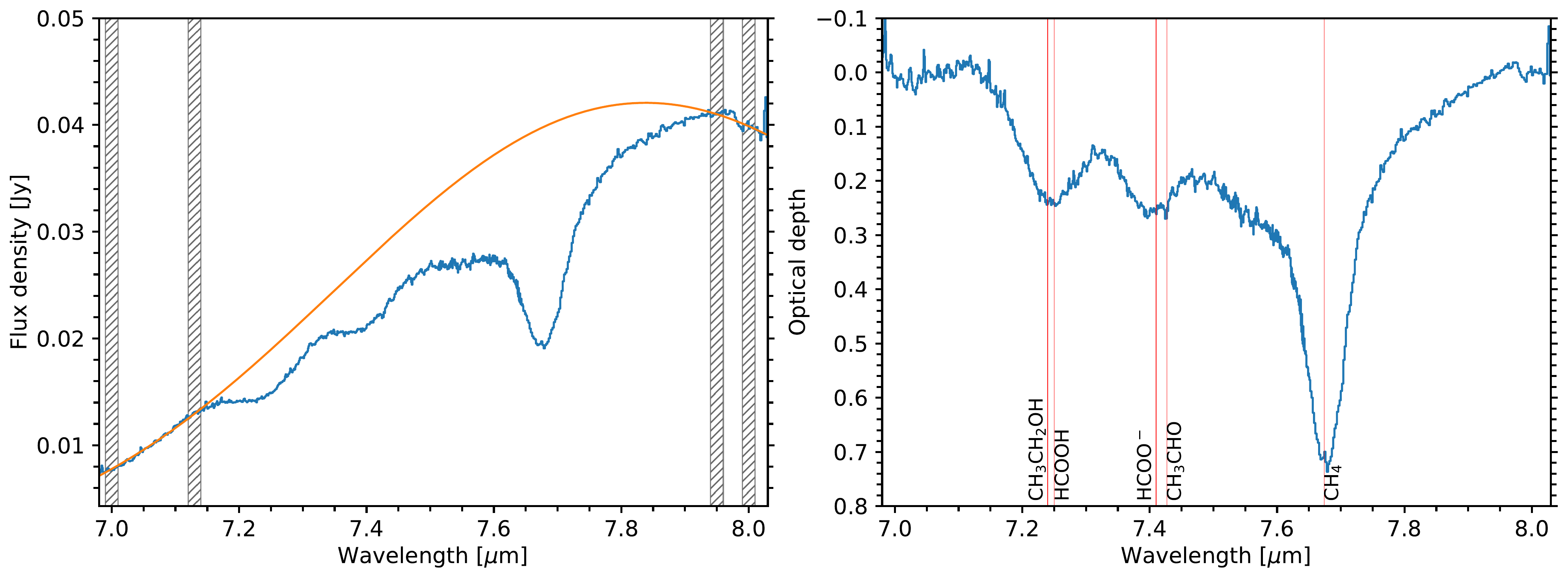}
  \caption{Ice features in the spectra from 7 to 8 \micron.  \textbf{Left}: The MRS spectra and the fitted continuum. The continuum is fitted locally with a third-order polynomial anchored at 6.99-7.01, 7.12--7.14, 7.94--7.96, and 7.99-8.01 \micron\ (gray hatched regions). \textbf{Right}: The optical depth spectra labeled with potential carriers.}
  \label{fig:ice_7-8}
\end{figure*}

This feature was tentatively detected in \source\ among a few other low-mass protostars, as well as high-mass protostars \citep{2008ApJ...678..985B,1999AA...343..966S}, but the low S/N of the optical depth spectra prohibited a robust carrier identification. We clearly detect the band at a high level of significance (Figure\,\ref{fig:ice_7-8}).  This feature could be associated with the CH$_3$ symmetric deformation mode of \ethanol\ \citep{2011ApJ...740..109O,2018AA...611A..35T} and/or the C--H/O--H deformation mode of HCOOH \citep{1999AA...343..966S,2007AA...470..749B}. The band strength of the HCOOH 7.24 \micron\ feature is $\sim$25 times weaker than that of its 5.83 \micron\ feature \citep{2007AA...470..749B}. Conversely, we estimate $\tau_{5.8\,\mu m}/\tau_{7.24\,\mu m}\sim 1.4$.  Despite considerable uncertainty in the fitted baseline and the \water\ absorption at 5.8 \micron, other species, such as \ethanol, may also contribute to the observed feature (Table\,\ref{tbl:com_ice_lab}).

\subsubsection{7.41 \micron\ feature: HCOO$^-$* and CH$_3$CHO*}
\label{sec:7.41}
This feature was tentatively seen in Spitzer/IRS spectra, but is clearly detected in the MIRI spectrum at high confidence. This feature may be due to the C=O stretching mode of HCOO$^-$ \citep{1999AA...343..966S} and/or the CH$_3$ symmetric deformation with the C--H wagging mode of CH$_3$CHO \citep{2011ApJ...740..109O,2018AA...611A..35T}. HCOO$^-$ has another C=O stretching mode at 6.33 \micron, where the observed spectrum has a slight bending feature at $\sim$6.31 \micron. CH$_3$CHO, on the other hand, has a feature at 7.427 \micron, located at a slightly longer wavelength than the observed feature. However, the peak position could move to 7.408 \micron\ depending on the ice mixture of CH$_3$CHO \citep{2018AA...611A..35T}. Thus, both species are potential contributors to this feature.

\subsubsection{7.7 \micron\ feature: CH$_4$} 
\label{sec:7.7}
This is a common feature attributed to the CH$_4$ deformation mode \citep{2008ApJ...678..985B}.  The optical depth of CH$_4$ is $\sim$0.6, while \citet{2008ApJ...678.1032O} measured a peak optical depth of 0.22$\pm$0.03 using Spitzer data.  The lower optical depth may be due to the much lower spectral resolving power ($R\sim 100$; $\Delta\lambda\sim 0.08\,$\micron) that under-resolves the narrow absorption feature (FWHM $\sim 0.07\,$\micron).  The higher spatial resolution in the MRS data may also result in a higher CH$_4$ optical depth, which varies spatially.
SO$_2$ ice has a feature at 7.63 \micron\ with a width of $\sim0.15$ \micron\ \citep{1997AA...317..929B}.  We cannot distinctively identify the contribution of SO$_2$ because of potential contribution from organic species, such as \ethanol\ (Table\,\ref{tbl:com_ice_lab}).

\subsubsection{9 \micron\ feature: NH$_3$} 
\label{sec:9}
\begin{figure*}[htbp!]
  \centering
  \includegraphics[width=\textwidth]{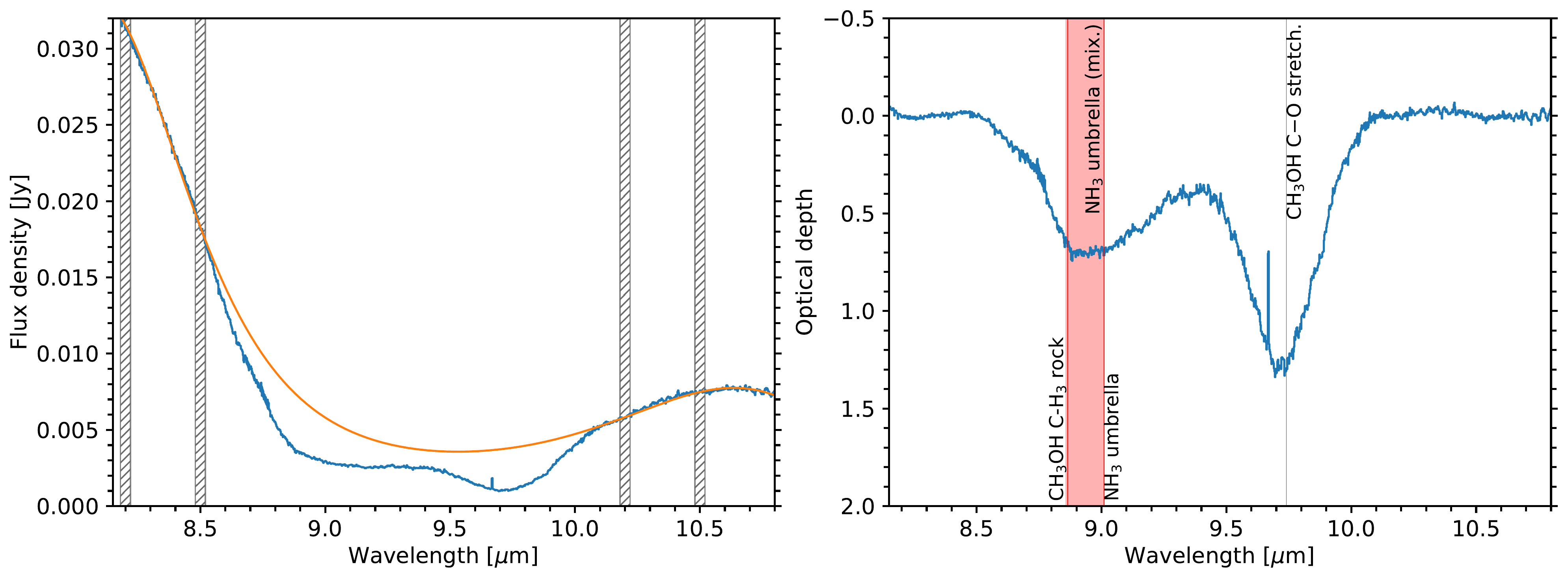}
  \caption{Ice features in the 8--11 \micron\ region of the spectrum. \textbf{Left}: The MRS spectrum and the continuum fit. The continuum is fitted locally with a fourth-order polynomial using 8.18--8.22, 8.48--8.52, 10.18--10.22, and 10.48--10.52 \micron\ as anchor points (gray hatched regions). \textbf{Right}: The optical depth spectrum labeled with potential carriers. The red shaded region highlights the range of the NH$_3$ umbrella mode in various ice mixtures investigated in \citet{2010ApJ...718.1100B}.}
  \label{fig:ice_8-11}
\end{figure*}

Both the CH$_3$ rocking mode of \methanol\ at 8.87 \micron\ and the umbrella mode of NH$_3$ at 9.01 \micron\ are likely to contribute to this feature (Figure\,\ref{fig:ice_8-11}). The former feature is narrower (FWHM=0.24 \micron) than the latter (FWHM=0.58 \micron). 
\citet{2010ApJ...718.1100B} showed that the peak position of the NH$_3$ umbrella mode could shift toward shorter wavelengths when mixed with H$_2$O and/or \methanol. \ethanol\ has its CH$_3$ rocking mode at 9.17 \micron\ and C--O stretching mode at 9.51 \micron. However, both features are very narrow (FWHM$\sim$0.1--0.2 \micron), and are not clearly visible in the MIRI spectra.

\subsubsection{9.7 \micron\ feature: \methanol} 
\label{sec:9.7}
This feature is commonly attributed to the C--O stretching mode of \methanol\ at 9.74 \micron. While the peak and width of the observed feature matches the expected \methanol\ absorption feature, there is slightly more absorption at the shorter wavelength side of the feature, hinting at contribution from other species, such as NH$_3$ and \ethanol\  (Section\,\ref{sec:composite}). A model of the silicate band, taking into account grain composition and size distribution, is required to accurately extract the profiles of the ice bands in this region, which is beyond the scope of this overview paper.

\subsubsection{11 \micron\ feature: \water\ libration} 
\label{sec:11}
\begin{figure*}[htbp!]
  \centering
  \includegraphics[width=\textwidth]{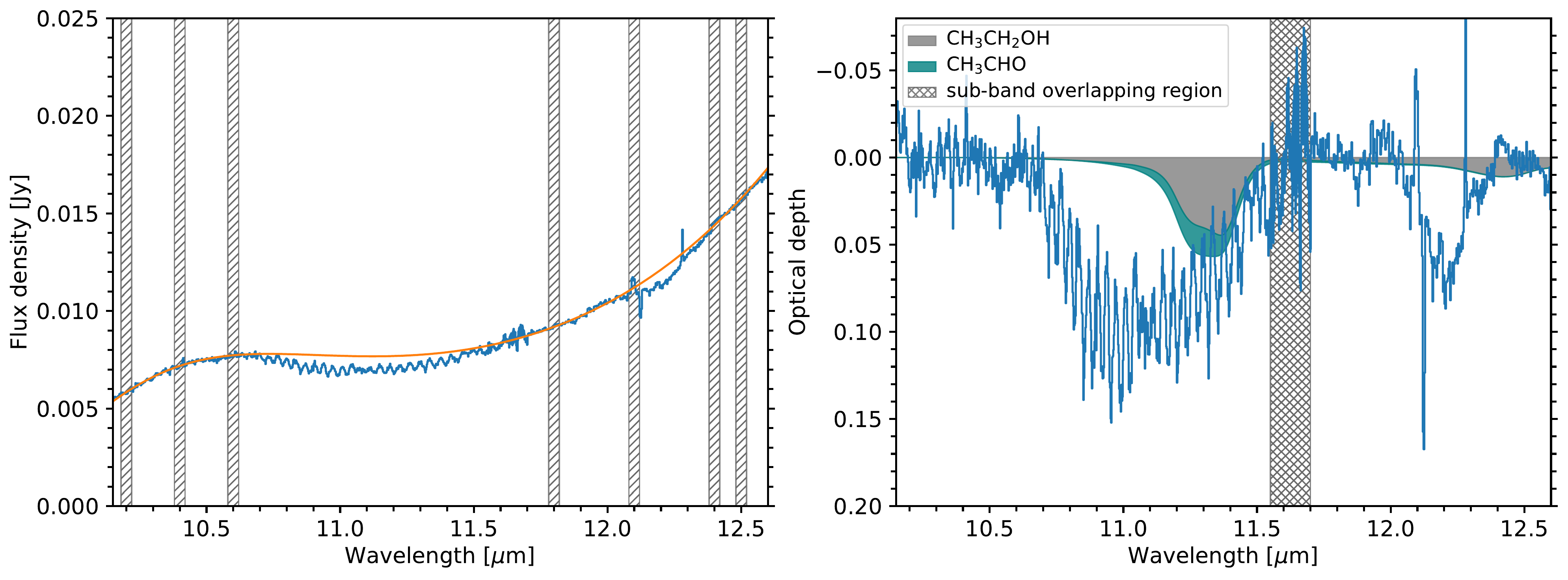}
  \caption{Ice features in the 11--12 \micron\ region of the spectrum. \textbf{Left}: The MRS spectrum and the continuum fit. The continuum is fitted locally with a fifth-order polynomial using $\pm$0.01 \micron\ regions around 10.18--10.22, 10.38--10.42, 10.58--10.62, 11.78--11.82, 12.18--12.22, 12.38--12.42, and 12.48--12.52 \micron\ as anchor points (gray hatched regions).  \textbf{Right}: The optical depth spectra compared to laboratory spectra of \ethanol, \acetaldehyde, and \methylformate.  The cross-hatched region highlights the overlapping range of two sub-bands, which has less stable baseline.}
  \label{fig:ice_11-12}
\end{figure*}

This feature is very broad, spanning 10--13 \micron, consistent with the well-known \water\ libration mode, which can extend to 30 \micron. \citet{2000ApJ...544L..75B} reported a narrower, weak absorption feature at 11.2 \micron, interpreted as polycyclic aromatic hydrocarbon (PAH) mixtures. Crystalline silicates, especially forsterite, also have absorption features around $\sim$11\,\micron\ \citep{2005ApJ...622..404K,2016MNRAS.457.1593W,2020MNRAS.493.4463D}. Finally, \citet{2021AA...651A..95T} showed that \ethanol, \acetaldehyde, and \methylformate\ could produce absorption at similar wavelengths. Figure\,\ref{fig:ice_11-12} shows the presence of an unambiguous 11.2\,\micron\ feature in the MIRI spectrum. Determining the carrier of this feature would require additional modeling.

\subsubsection{15.2 \micron\ CO$_2$} 
\label{sec:15.2}
\begin{figure*}[htbp!]
  \centering
  \includegraphics[width=\textwidth]{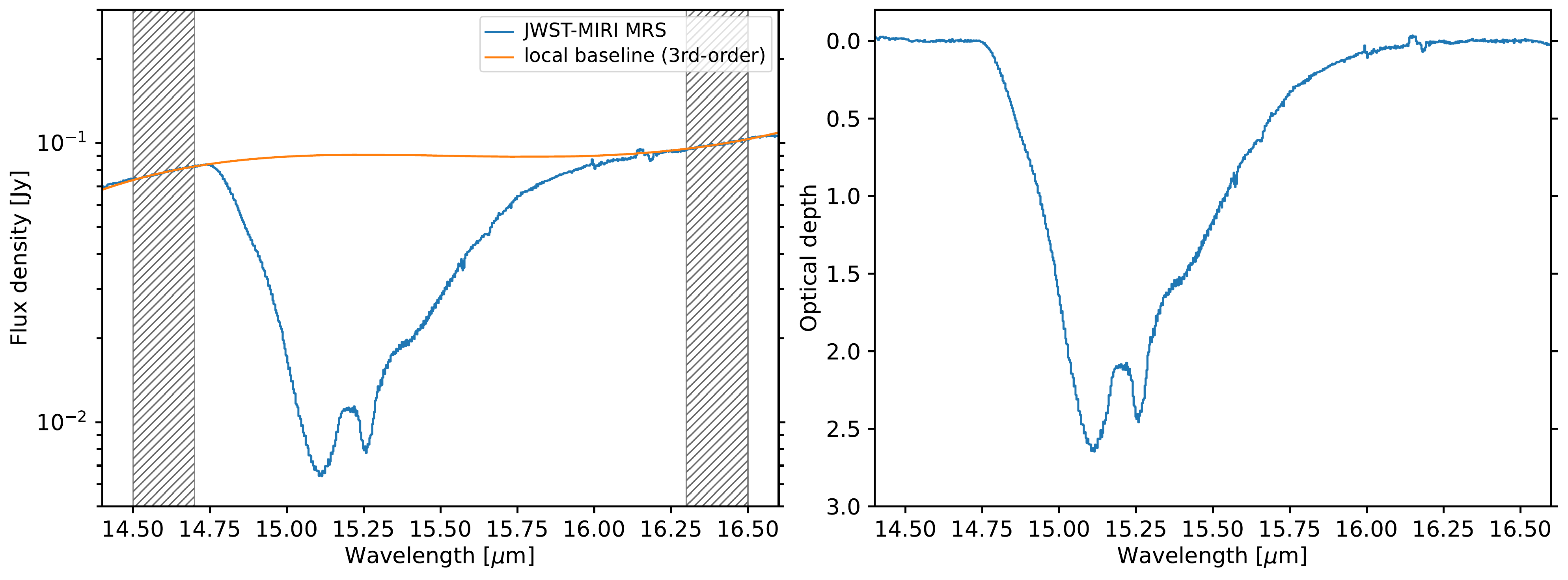}
  \caption{Spectra of the CO$_2$ ice feature at 15.2 \micron.  The MIRI spectra zoomed to the CO$_2$ ice features are shown on the left.  A local baseline (orange) is fitted with a third-order polynomial using the spectra in 14.5--14.7 \micron\ and 16.3--16.5 \micron.  The derived optical depth spectra are shown on the right.}
  \label{fig:co2}
\end{figure*}

This ubiquitous feature is due to the bending mode of CO$_2$ (Figure\,\ref{fig:co2}). The double peaks are a distinctive signature of crystalline, usually relatively pure, CO$_2$ ice \citep{1997AA...328..649E}. 
There are two broader features at 15.1 and 15.3 \micron, corresponding to the apolar CO$_2$:CO mixture and the polar CO$_2$:\water\ mixture, respectively.  The shoulder extending toward longer wavelengths is due to CO$_2$ mixed with \methanol.  \citet{2008ApJ...678.1005P} detected the double-peaked CO$_2$ with Spitzer in the same source; however, the strength of those peaks was weaker than the MRS spectra indicate.  The significantly improved spectral resolution may lead to stronger peaks, but constraining the origin of such change, such as a temporal variation, requires further modeling.

Pure CO$_2$ ice only form in regions with elevated temperature, at $\sim50-80$ K via the thermal annealing process \citep{1999ApJ...522..357G,2013PNAS..11012899E,2018ApJ...869...41H} or at $\sim20-30$ K via the distillation of a CO$_2$:CO mixture \citep{2008ApJ...678.1005P}.  \citet{2011ApJ...729...84K} suggest that detection of pure CO$_2$ in low-luminosity protostars could be indicative of previous episodic accretion.  In fact, \citet{2013ApJ...779L..22J} found a ring-like (inner radius of 150--200 au) structure of H$^{13}$CO$^+$ emission with ALMA, suggesting that water vapor is present on small scales destroying H$^{13}$CO$^+$ \citep{1992ApJ...399..533P}.  The origin of this water vapor could be an accretion burst that occurred 100--1000 years ago, increasing the luminosity by a factor of 100, making such an interpretation for the CO$_2$ double peak a viable explanation.  In the distillation scenario, both a warm disk and the inner envelope can provide suitable environments; however a well-defined Keplerian disk has not yet been detected in \source.

\subsection{Composite ice spectra}
\label{sec:composite}
\begin{figure*}[htbp!]
    \centering
    \includegraphics[width=\textwidth]{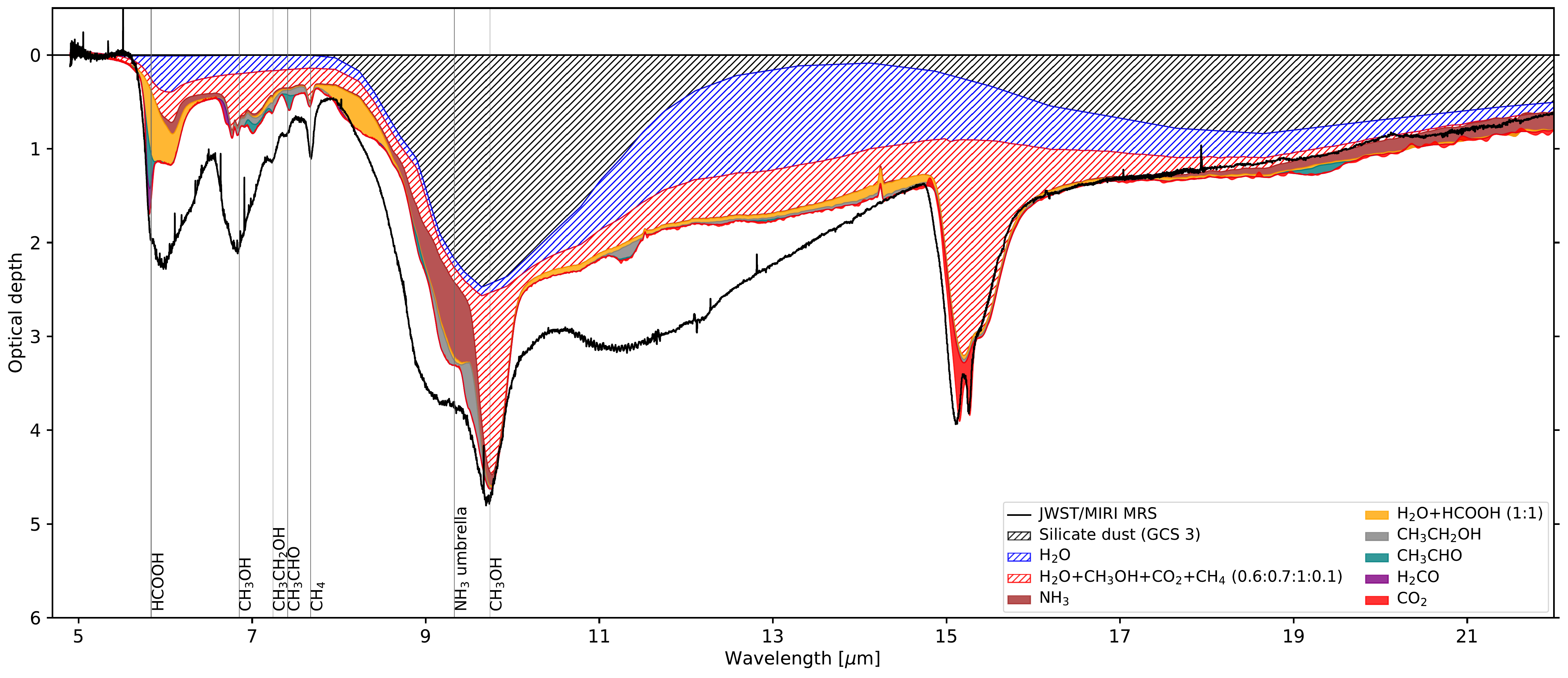}
    \caption{Optical depth spectra compared with a composite spectra of several ice species measured from experiments.  The details of the laboratory data are discussed in Appendix\,\ref{sec:lab_data}.  The ``CO$_2$'' spectra (red) are from pure CO$_2$.
    }
    \label{fig:tau_lab_comp}
\end{figure*}

The unprecedented S/N combined with the sub-arcsec spatial resolution allows a multi-component ice spectral comparison with laboratory data across the entire range of MIRI coverage (4.9--28 \micron).  As discussed in Section\,\ref{sec:features}, many absorption features are likely to have several contributing ice species, and only the strongest features could be robustly identified by previous studies.  The highly sensitive MIRI MRS spectrum enables a comprehensive approach to compare composite optical depth spectra including multiple ice species.  Figure\,\ref{fig:tau_lab_comp} shows a simple composite synthetic spectrum of several ice species discussed in Section\,\ref{sec:features}.  We also include the spectrum of GCS 3, representing the silicate dust \citep{2004ApJ...609..826K}.   The optical depth spectrum of each ice species and mixture is scaled to match the observations.  While we do not aim to fit the observed optical depth spectra, we can already see wavelength regions where the laboratory ice spectra reproduce the observations in this toy model, such as $\sim$10 \micron\ and $\sim$15 \micron.  This simple model underestimates the absorption at 5--9 \micron\ and 11--12 \micron\ regions, calling for detailed ice modeling in future studies.  This experiment demonstrates the vast potential of JWST/MIRI spectroscopy for studies of interstellar ices.

\section{Warm Water Vapor and CO Gas as a Signpost of the Embedded Disk}
\label{sec:water_vapor}

\begin{figure*}[htbp!]
  \centering
    \includegraphics[width=18cm]{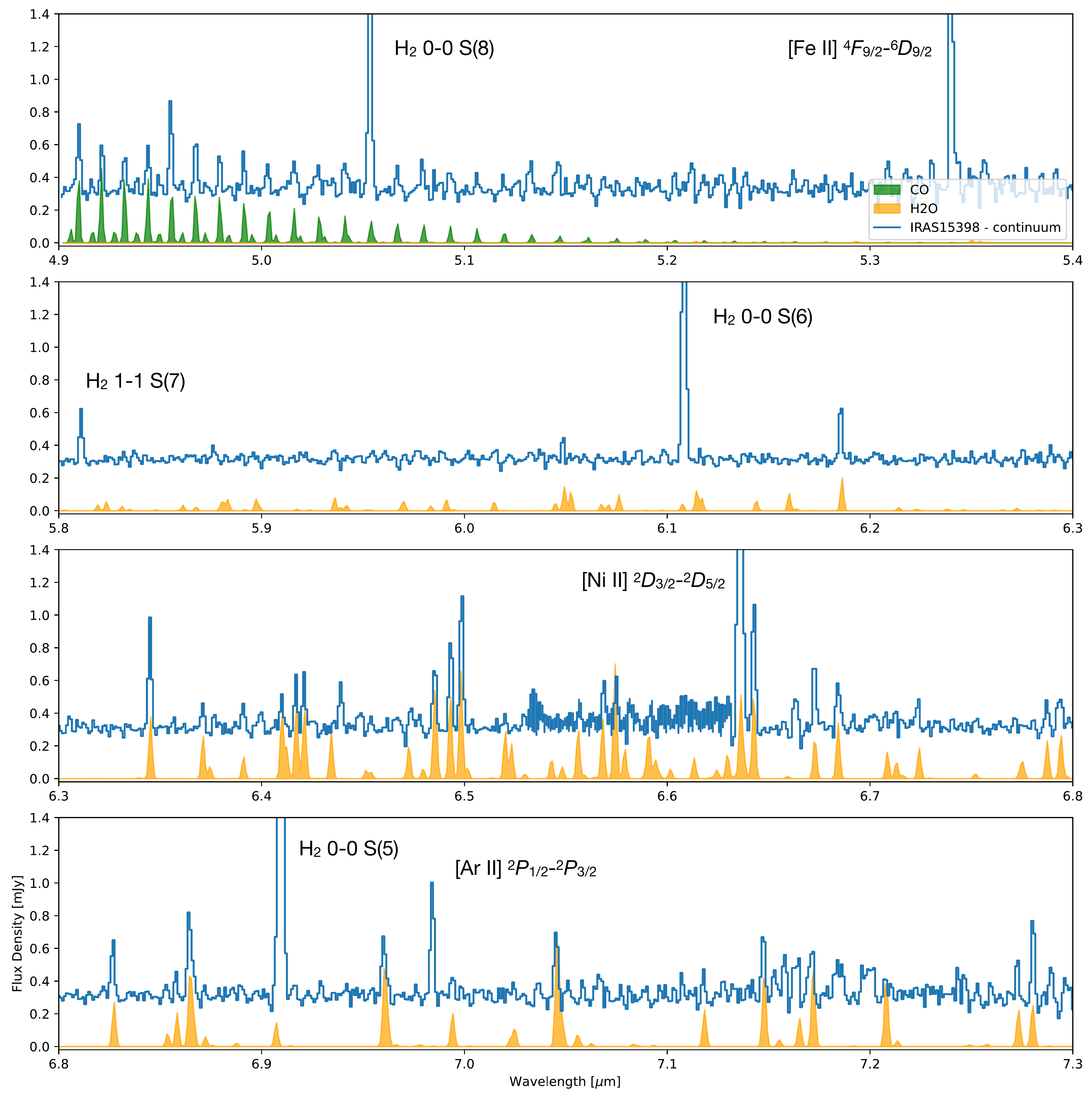}
\caption{Continuum-subtracted spectra showing emission from the CO fundamental P-branch and the H$_2$O bending mode centered on 6\,$\mu$m. The data, which is shown with an offset of 0.3 mJy, are modeled by a simple slab model for line identification \citep{2011ApJ...743..112S,2020zndo...4037306S}.  Emission lines other than \water\ and CO are annotated.}
  \label{fig:h2o_co}
\end{figure*}

JWST provides spatial resolution similar to that achieved by ALMA, allowing us to search for signatures of the embedded disk suggested by ALMA observations \citep{2017ApJ...834..178Y,2018ApJ...864L..25O}.  Warm water and CO gas at $M$-band (4.7--5 \micron) are a common tracer of the inner disk in Class I and II sources \citep{2003AA...408..981P,2022AJ....163..174B}, but they have rarely been detected in Class 0 sources, like \source.  In Figure\,\ref{fig:h2o_co}, we compare the baseline-subtracted 4.9--7.3 $\mu$m region of the \source\ spectrum with a simple slab model of warm water vapor ($\sim200-300$ K) and CO fundamental ($\nu=1-0$ and $2-1$) ro-vibrational lines at a higher temperature \citep{2011ApJ...743..112S,2020zndo...4037306S}.  The synthetic spectra are multiplied with the continuum to account for variable extinction on these emission lines, which fit the data better. The molecular data are taken from HITRAN \citep{GORDON2022107949}.  The water lines appear prominently from 5.8--7.3 $\mu$m, while the (P-branch; $\Delta J = -1$) CO appears at the shortest MIRI wavelengths (4.9--5.3 $\mu$m). Although these models are not adapted to this source, it is clear from inspection that the region contains a large number of compact emission lines. 

The agreement between model and observation is considerable. We can state with confidence that the majority of this emission comes from a compact region of the source, and is attributable to warm water vapor, which is likely excited in the previously undetected embedded disk region, within the inner 0.2$\arcsec$, and/or the shocked gas in the inner envelope. The specific model fits and constraints on the spatial extent of the emission are left to a future work.

\section{Outflows and Jets}
\label{sec:outflows}
\subsection{MIRI Imaging}
\label{sec:imaging}
\begin{figure*}[htbp!]
  \centering
  \includegraphics[width=\textwidth]{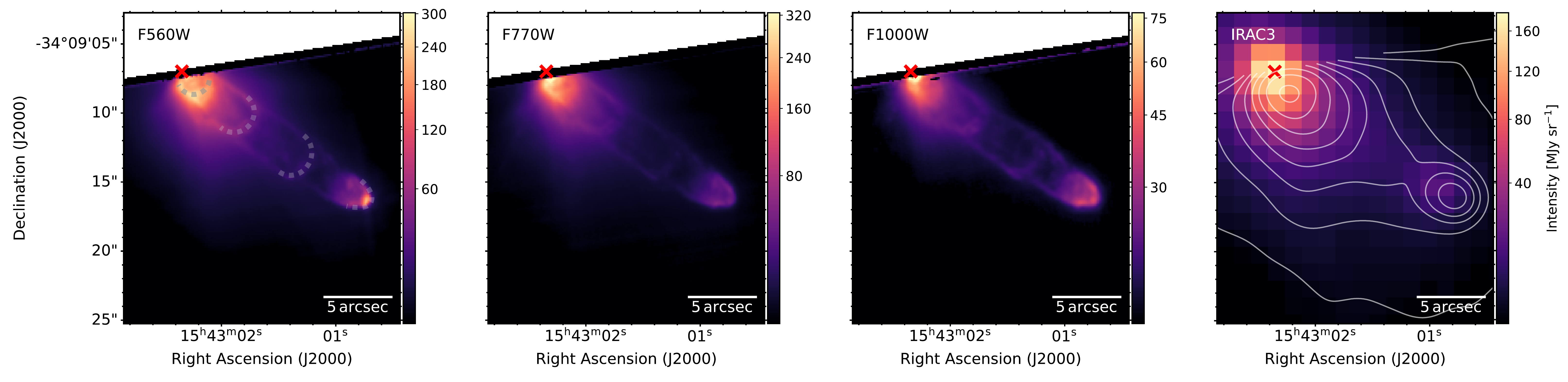}
  \caption{MIRI images with F560W, F770W, and F1000W filter taken in parallel with the background pointing.  The identified four shell-like structures are highlights in the panel of the F560W image (left). The image on the right shows the archival Spitzer IRAC 3 image taken on 2004 September 3, where the MIRI F560W image convolved with the IRAC 3 resolution of 1.88\arcsec\ is shown in logarithmic contours from 1 to 150 MJy sr$^{-1}$.  The FoV of the MIRI imager leaves a distinct shape at the top left of the images.  The red ``x'' labels the sub-mm continuum peak position.}
  \label{fig:miri_image}
\end{figure*}

The parallel imaging of our background pointing serendipitously covered the blue-shifted outflow of \source.  Figure\,\ref{fig:miri_image} shows the MIRI images of the blue-shifted outflows in three filters.  The F560W image contains both the continuum and the H$_2$ S(7) line; the F770W image includes the continuum and the H$_2$ S(4) line; and the F1100W image consists of the continuum and the H$_2$ S(3) line.  These images unveil exquisite details in the outflow, showing at least four shell-like structures.  The outermost shell appears similar to a terminal bow-shock.  The opening angle of each shell decreases with the distance from the protostar.  ALMA observations of outflow tracers, such as CO, H$_2$CO, and CS, show similar shell-like variations \citep{2016AA...587A.145B,2020ApJ...900...40O,2021ApJ...910...11O}, for which \citet{2021AA...648A..41V} interpret as precessing episodic outflows driven by a jet. Compared to archival IRAC images taken in 2004 September 3, the terminal shock knot moved by $1.8\arcsec$ along the outflow, which is measured from the centroids of the fitted 2D Gaussian profiles to the blob in the IRAC 3 image and the MIRI F560W image convolved with the IRAC 3 resolution of 1.88\arcsec\ (Figure\,\ref{fig:miri_image}).  Considering a length of $\sim17\arcsec$ measured in our MIRI images, the dynamical time of the blue-shifted outflow is, thus, $\sim$170 years, suggesting an extremely recent ejection.  \citet{2021AA...648A..41V} also identified four ejections separated by 50--80 years.  Interestingly, the mid-IR outflow has almost the same morphology as the molecular outflow observed in sub-mm.

\subsection{Spectral Line Emission}
\label{sec:emission}

\begin{deluxetable}{ccc}
  \tablecaption{Detected Emission Lines}
  \label{tbl:emission_lines}
  \tablehead{
  \colhead{Wavelength} & \colhead{Species} & \colhead{Transition} \\
  ($\mu$m) & & 
  } 
  \startdata
  5.053   & H$_2$             & 0--0 S(8) \\
  5.340   & [Fe\,\textsc{ii}] &  $^4F_{9/2}$--$^6D_{9/2}$ \\
  5.511   & H$_2$             & 0--0 S(7) \\
  5.811   & H$_2$             & 1--1 S(7) \\
  6.109   & H$_2$             & 0--0 S(6) \\
  6.636   & [Ni\,\textsc{ii}] & $^2D_{3/2}$--$^2D_{5/2}$ \\
  6.910   & H$_2$             & 0--0 S(5) \\
  6.985   & [Ar\,\textsc{ii}] & $^2P_{1/2}$--$^2P_{3/2}$ \\
  8.025   & H$_2$             & 0--0 S(4) \\
  9.665   & H$_2$             & 0--0 S(3) \\
  12.279  & H$_2$             & 0--0 S(2) \\
  12.814  & [Ne\,\textsc{ii}] &  $^2P^0_{1/2}$--$^2P^0_{3/2}$ \\
  17.035  & H$_2$             & 0--0 S(1) \\ 
  17.936  & [Fe\,\textsc{ii}] &  $^4F_{7/2}$--$^4F_{9/2}$ \\
  24.519  & [Fe\,\textsc{ii}] &  $^4F_{5/2}$--$^4F_{7/2}$ \\
  25.249  & [S\,\textsc{i}]   &  $^3P_1$--$^3P_2$ \\
  25.988  & [Fe\,\textsc{ii}] &  $^6D_{7/2}$--$^6D_{9/2}$ \\
  \enddata
\end{deluxetable}

\begin{figure*}[htbp!]
  \centering
  \includegraphics[width=\textwidth]{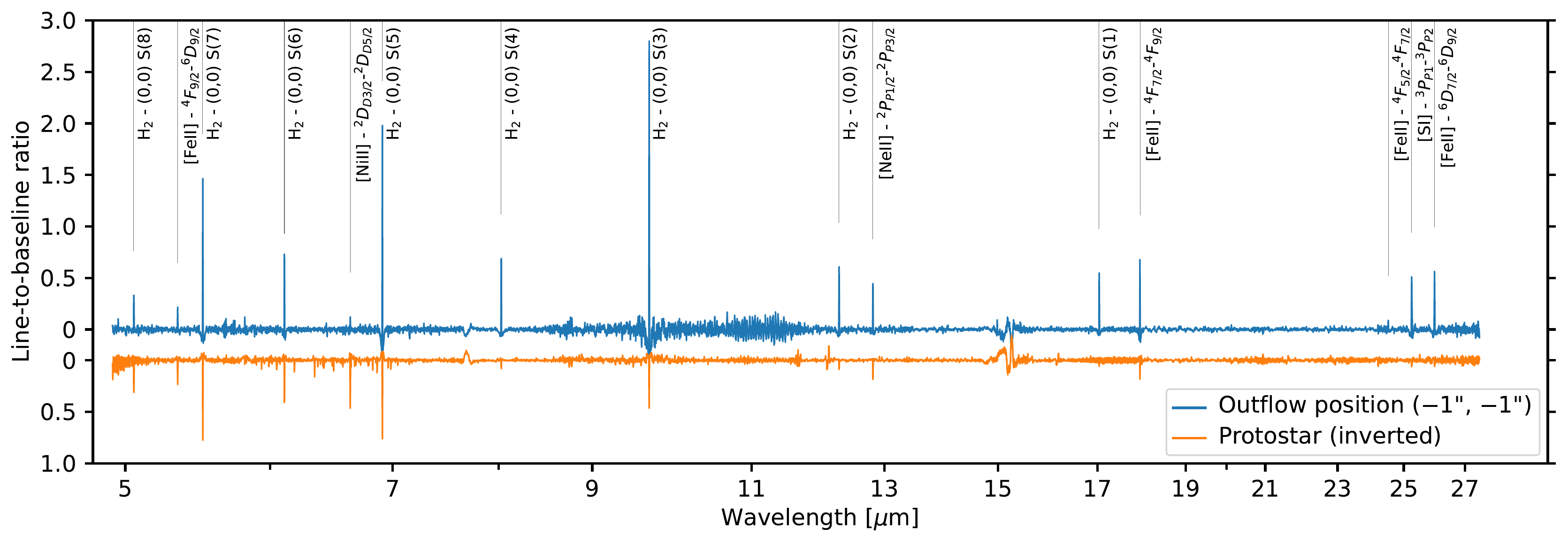}
  \caption{The baseline-divided spectrum extracted on-source (orange) compared with the outflow spectrum extracted at ($-$1\arcsec, $-$1\arcsec) from the sub-mm continuum peak (blue).  Both spectra are extracted with a fixed 1\arcsec\ aperture.  The on-source spectrum is multiplied by $-1$ for better visual comparison with the outflow spectrum.  The weaker emission lines at 5--7 \micron\ in the on-source spectrum are mostly \water\ and CO discussed in Section\,\ref{sec:water_vapor}; and the noise in the $\sim$9--11 \micron\ range of both spectra is due to the fringe residual.  The continuum is derived by convolving a Gaussian profile with a width of 10 channels.  Imperfect continuum subtraction appears around major ice features, such as $\sim$7 and $\sim$15 \micron.  Identified emission lines are annotated. }
  \label{fig:emission_lines}
\end{figure*}

We also identified several emission lines in the MRS spectra besides the CO and \water\ lines.  We extracted a 1D spectrum at ($15^{\mathrm{h}}43^{\mathrm{m}}02.16^{\mathrm{s}}$ $-34^\circ09{}^\prime07.99{}^{\prime\prime}$), which is ($-$1\arcsec, $-$1\arcsec) from the sub-mm continuum peak, with an aperture of 1\arcsec\ to better probe the emission due to outflow activity (Figure\,\ref{fig:emission_lines}). Most lines appear strong in outflows compared to the spectrum toward the protostar, except for the [Ni\,\textsc{ii}] line at 6.636 \micron. Veiling due to scatter light and extinction from the envelope are not considered in this simple extraction, which aims to present a qualitative view of the detected emission lines.  As noted in Table \ref{tbl:emission_lines}, most of the strong line emission is identified with either H$_2$ pure rotational lines or ionized/neutral fine-structure lines from Fe, Ne, or S.
Previously with Spitzer IRS spectra, \citet{2010AA...519A...3L} detected H$_2$, S(1) and S(4), [Fe\,\textsc{ii}], 17.9 and 26.0 \micron, as well as the [Si\,\textsc{ii}] 35 \micron\ in \source, the last of which is not covered by MIRI. All of these lines are spatially extended in a bipolar pattern on the NW-SE axis.  There is tentative evidence of other weaker emission from the species mentioned in Table\,\ref{tbl:emission_lines}.  We defer a comprehensive analysis of emission lines to a future paper.

\begin{figure*}[htbp!]
  \centering
  \includegraphics[width=\textwidth]{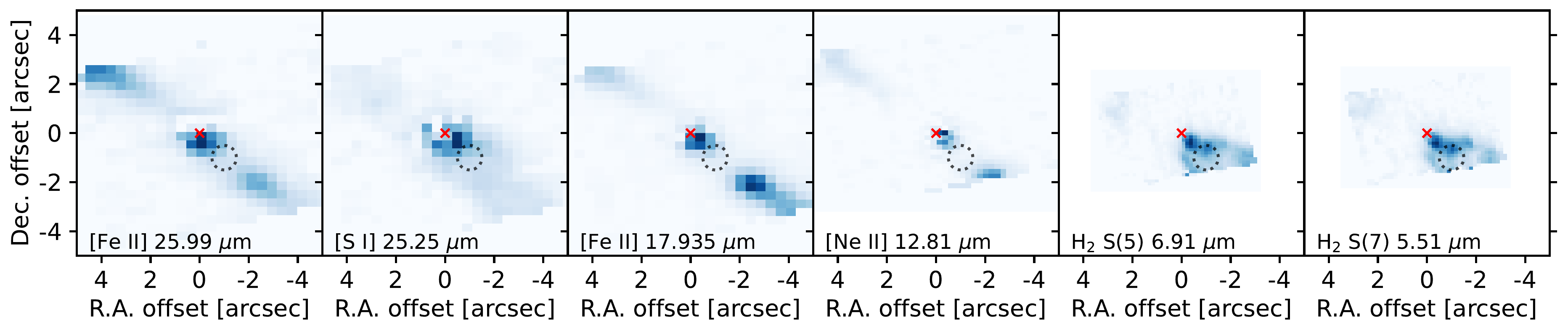}
  \caption{Continuum-subtracted intensity maps of [Fe\,\textsc{ii}] lines at 25.988 and 17.935 \micron, [S\,\textsc{i}] at 25.249 \micron, [Ne\,\textsc{ii}] at 12.814 \micron, H$_2$ S(5) and S (7) at 6.910 and 5.511 \micron, respectively.  The map is calculated from the average intensity within $\pm$200 \kms\ from the rest frame line centroids.  The continuum is calculated from $-$500 to $-$300 \kms.  The red ``x'' labels the sub-mm continuum peak position.  The dotted circles indicate the extraction aperture for the spectrum shown in Figure\,\ref{fig:emission_lines}.}
  \label{fig:emission_maps}
\end{figure*}

Figure\,\ref{fig:emission_maps} shows the continuum-subtracted intensity maps of several representative ionic and molecular lines. The molecular lines, such as H$_2$, show a broad opening angle morphology and appear to highlight the walls of the shocked cavity.  They also show sub-structures mostly within the south-western (blue-shifted) outflow cavity.  The ionic lines, such as [Fe\,\textsc{ii}] and [Ne\,\textsc{ii}], likely represent hotter regions and are tightly collimated into a jet within the cavity region. In most cases, the ionic lines are spectrally resolved across a few channels, corresponding to a velocity range of $\pm$200 \kms. The ionic lines are generally associated with outflows and connected to accretion processes in the central protostar \citep{2016ApJ...828...52W}. 

\section{Conclusions}
\label{sec:conclusions}
It is clear from these first observations of \source\ that JWST MIRI will transform our understanding of protostellar ice chemistry as well as ice chemistry in all environments. We present detections of previously identified ice species and provide evidence for the possible presence of organic ice species.  We also show gaseous emission of warm water and CO, which is often found in warm disks.  Other detected emission lines, including H$_2$, [Fe\,\textsc{ii}], [Ne\,\textsc{ii}], and [S\,\textsc{i}], appear extended along the outflow direction, tracing a wide-angle outflow cavity and a collimated jet.  The MIRI imaging serendipitously captured the south-western outflow of \source, providing us an exquisite view of the outflow structure in the infrared.  

The main conclusions of this first analysis of the JWST/MIRI observations of \source\ are summarized below.

\begin{itemize}
    \item A MIRI MRS spectrum of a Class 0 protostar, \source, is reported for the first time.  The protostar appears as a point source over the full wavelength range at 5--28 \micron.
    \item The MRS data show rich ice absorption features.  Particularly, the ice features between 5 and 8 \micron\ are detected with high S/N, allowing us to search for organic ice species.  We robustly identify ice species including \water, CO$_2$, CH$_4$, NH$_3$, \methanol, H$_2$CO, and HCOOH.  Furthermore, we detect ice absorption features that could imply the presence of NH$_4^+$, HCOO$^-$, \ethanol, \acetaldehyde, and \methylformate.  The CH$_4$ and pure CO$_2$ ice features appear stronger in the MIRI MRS spectra compared to previous Spitzer studies.  Significantly improved spectral resolution could result in deeper absorption, providing accurate constraints on the ice compositions.  Stronger absorption could also imply variability in ice column densities.
    \item The spectra between 5 and 8 \micron\ have many weaker emission lines.  The continuum-subtracted spectra present similar features to those from the synthetic spectra of warm water vapor and CO gas.  These emission lines only appear toward the protostar, hinting at warm water vapor and CO gas on small scales possibly on the disk surface.
    \item The MIRI imaging captures the blue-shifted outflow of \source, showing multiple shell-like structures consistent with the molecular outflows seen at sub-mm wavelengths.  The infrared outflow has similar length as the sub-mm outflow.  The proper motion of the compact shock knot indicates a dynamical time of $\sim$150 year for that ejection.
    \item Multiple emission lines are detected in the MRS spectra, including [Fe\,\textsc{ii}], [Ne\,\textsc{ii}], [S\,\textsc{i}], and H$_2$.  The H$_2$ S(8) line is the first detection in young protostars.
    \item The [Fe\,\textsc{ii}] and [Ne\,\textsc{ii}] emission show a collimated bipolar jet-like structure along the known outflow direction.  The emission also highlights a bright knot $\sim$2.5\arcsec\ away from the protostar toward southwest.  The emission of H$_2$ appears more extended, tracing a wide-angle outflow cavity.  
\end{itemize}

This JWST/MIRI observations of \source\ show striking details about solid-state features, providing the observational constraints for extensive searches of new ice species and detailed modeling of their abundances.  The characterization of gas-phase COMs has progressed significantly in the last decade, in large part due to the maturation of sub-mm interferometry (e.g., ALMA and NOrthern Extended Millimeter Array). Conversely, observational constraints on the ice-phase COMs are so far mostly from observations using ISO/SWS and Spitzer/IRS with limited spectral and spatial resolving power and sensitivity.  Absorption features of rare organic ice species in low mass protostars have low contrast and therefore require very high S/N and accurate spectro-photometric calibration to detect.  The absorption features between 7 and 8 $\mu$m were only detected in high-mass YSOs (e.g., W33A) with ISO-SWS, and similar features were only marginally detected with Spitzer in low-mass protostars. Consequently, the composition of organic ices around low-mass protostars has only been weakly constrained until now. With the advent of the JWST and the Mid Infrared Instrument (MIRI) spectrograph, the present observations definitively demonstrate that we can now detect and constrain mid-IR COM ice feature strength at high precision and provide much stronger guidance to models of gas-grain chemistry. 

\clearpage
\acknowledgements
This work is based on observations made with the NASA/ESA/CSA James Webb Space Telescope. The data were obtained from the Mikulski Archive for Space Telescopes at the Space Telescope Science Institute, which is operated by the Association of Universities for Research in Astronomy, Inc., under NASA contract NAS 5-03127 for JWST. These observations are associated with JWST GO Cycle 1 program ID 2151. Y.-L. Yang acknowledges support from the Virginia Initiative of Cosmic Origins Postdoctoral Fellowship.  Y.-L. Yang and N. Sakai acknowledge support from a Grant-in-Aid from the Ministry of Education, Culture, Sports, Science, and Technology of Japan (20H05845, 20H05844), and a pioneering project in RIKEN (Evolution of Matter in the Universe). L.I.C. acknowledges support from the David and Lucille Packard Foundation, Johnson \& Johnson WISTEM2D, and NASA ATP 80NSSC20K0529.  Y.-L. Yang thanks J. Terwisscha van Scheltinga for laboratory ice spectra, Y. Okoda for useful discussion on the ALMA observations of the presented source, and S. Zeng and R. Nakatani for the motivation to explore the MIRI imaging products. L. I. Cleeves, R. T. Garrod, B. Shope, J. B. Bergner, C. N. Shingledecker, K. M. Pontoppidan, and J. D. Green acknowledges support from NASA/STScI GO grant JWST-GO-02151. J.-E. Lee and C.-H. Kim were supported by the National Research Foundation of Korea (NRF) grant funded by the Korea government (MSIT) (grant number 2021R1A2C1011718).  EvD is supported by EU A-ERC grant 101019751 MOLDISK and by the Danish National Research Foundation (grant agreement no. DNRF150, ``InterCat''). This research benefited from discussions held with the international team \#461 ``Provenances of our Solar System's Relics'' (team leaders Maria N. Drozdovskaya and Cyrielle Opitom) at the International Space Science Institute, Bern, Switzerland. This research has made use of the NASA/IPAC Infrared Science Archive, which is funded by the National Aeronautics and Space Administration and operated by the California Institute of Technology. The National Radio Astronomy Observatory is a facility of the National Science Foundation operated under cooperative agreement by Associated Universities, Inc.  This research made use of Photutils, an Astropy package for detection and photometry of astronomical sources \citep{larry_bradley_2022_6825092}.  This research made use of APLpy, an open-source plotting package for Python \citep{aplpy2012,aplpy2019}.

\facilities{JWST, Spitzer, IRSA}

The JWST data used in this paper can be found in MAST: \dataset[10.17909/wv1n-rf97]{\doi{10.17909/wv1n-rf97}}.

\software{astropy v5.1 \citep{2013AA...558A..33A,2018AJ....156..123A}, jwst \citep{2019ASPC..523..543B}, photutils v1.5.0 \citep{larry_bradley_2022_6825092}, aplpy v2.1.0 \citep{aplpy2019}}

\appendix

\section{Characteristics of the Extraction Apertures}
\label{sec:extraction}
\begin{figure}[htbp!]
  \centering
  \includegraphics[width=\textwidth]{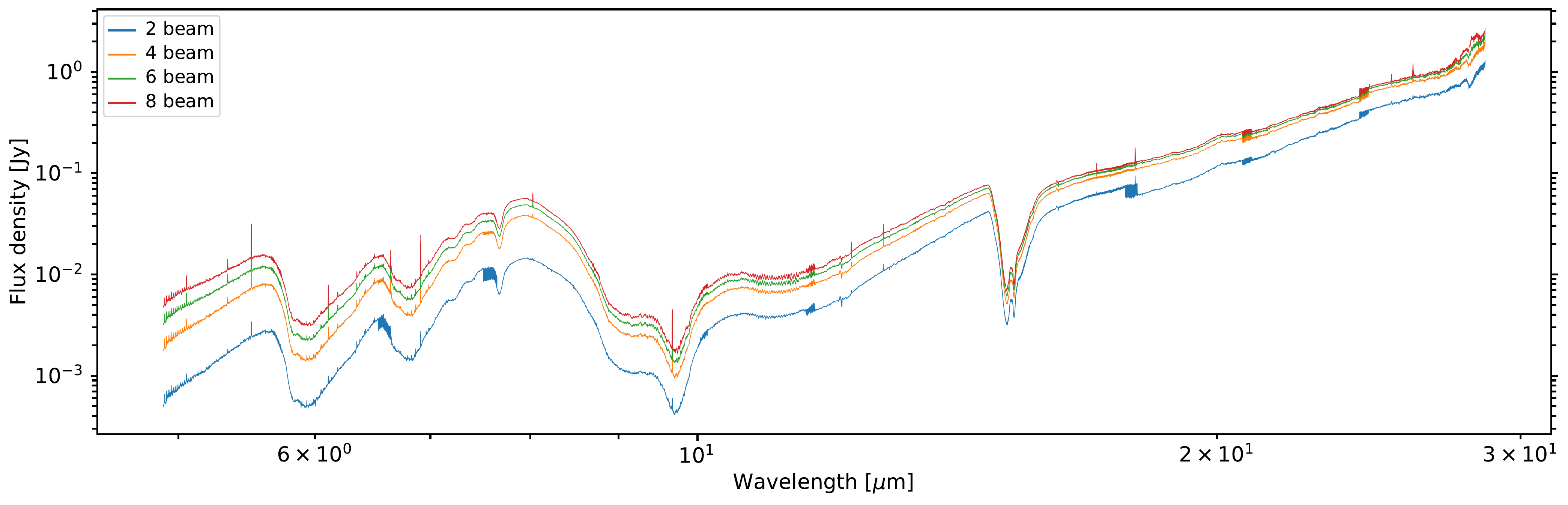}
  \includegraphics[width=\textwidth]{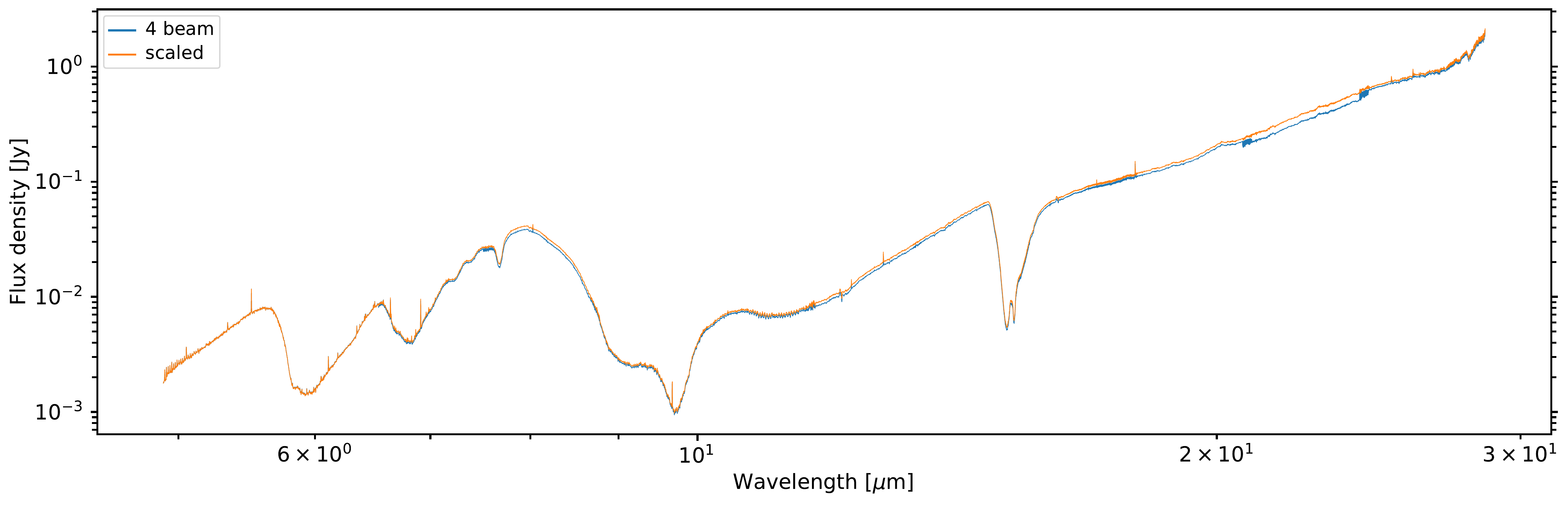}
  \caption{\textbf{Top:} The 1D spectra extracted with aperture sizes of 2, 4, 6, and 8 beams without scaling.  \textbf{Bottom:} The 1D spectra extracted with a 4-beam aperture with and without scaling.}
  \label{fig:extraction}
\end{figure}

The protostar appears point-like in the MRS spectral cube, showing the Airy pattern most noticeably at the longer wavelengths.  Therefore, to extract 1D spectra, we define the aperture in units of the diffraction-limited beam, resulting in variable aperture increasing with wavelength.  Because the source is not a perfect point source, we expect the 1D spectrum extracted with a small aperture would lead to missing flux if the emission is more extended due to scattering; the actual beam size may be larger than the diffraction-limited beam size due to the detector scattering at shorter wavelength.  On the other hand, a larger aperture may start to add noise to the 1D spectrum.  
The extracted 1D spectra with different aperture sizes demonstrate the aforementioned effects (Figure\,\ref{fig:extraction}, top).  The 4-beam aperture extraction results in a good balance between missing flux and noise, which is adopted in this study for extracting the 1D spectrum.  The spectrum extracted with a 4-beam aperture with the median scaling between sub-bands (see Section\,\ref{sec:observations}) differs from the un-scaled spectrum by up to 16\% (Figure\,\ref{fig:extraction}, bottom).

\section{Comparison between JWST/MIRI MRS Spectra and Spitzer/IRS Spectra}
\label{sec:irs_comp}
\begin{figure}[htbp!]
  \centering
  \includegraphics[height=2.3in]{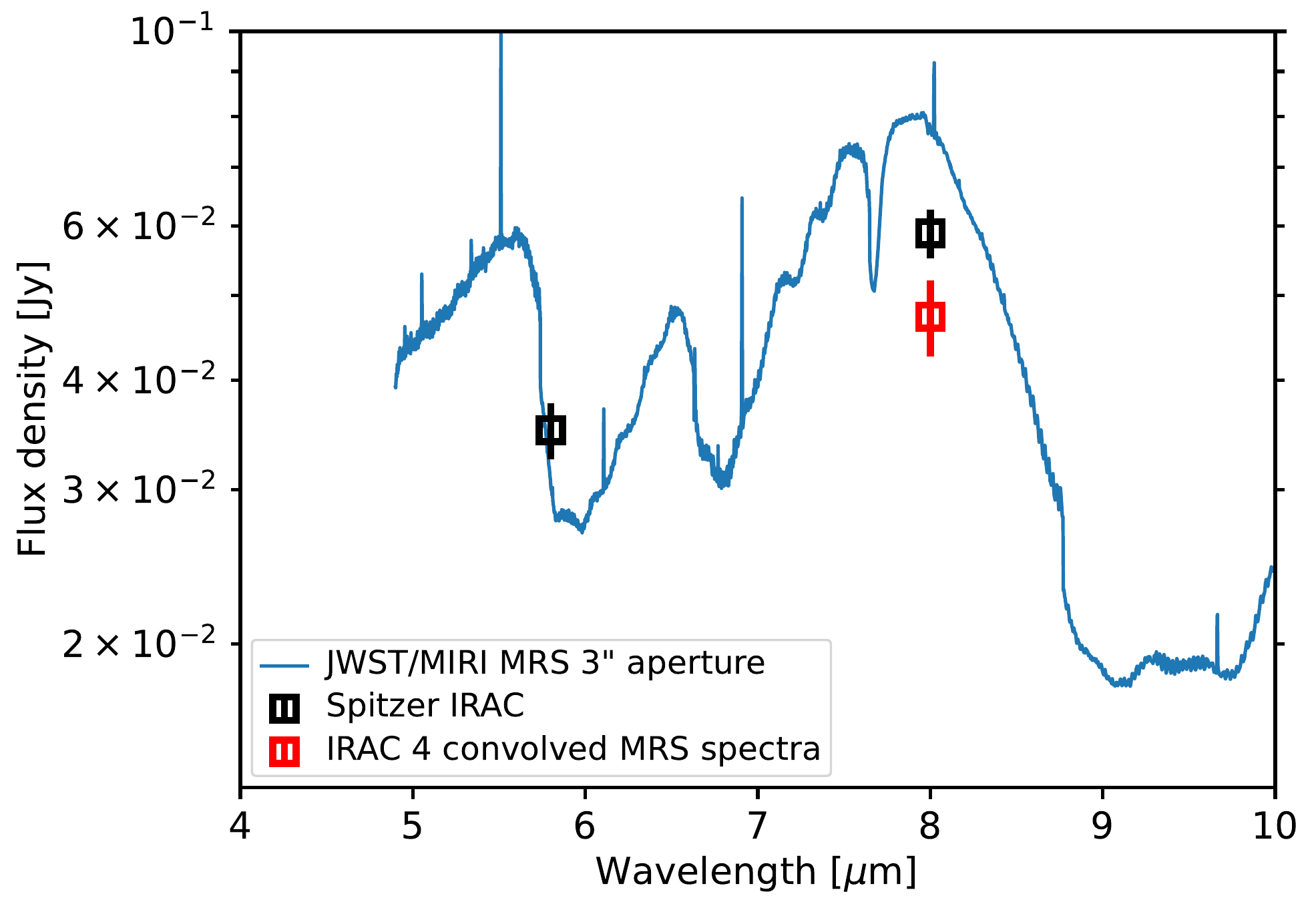}
  \includegraphics[height=2.3in]{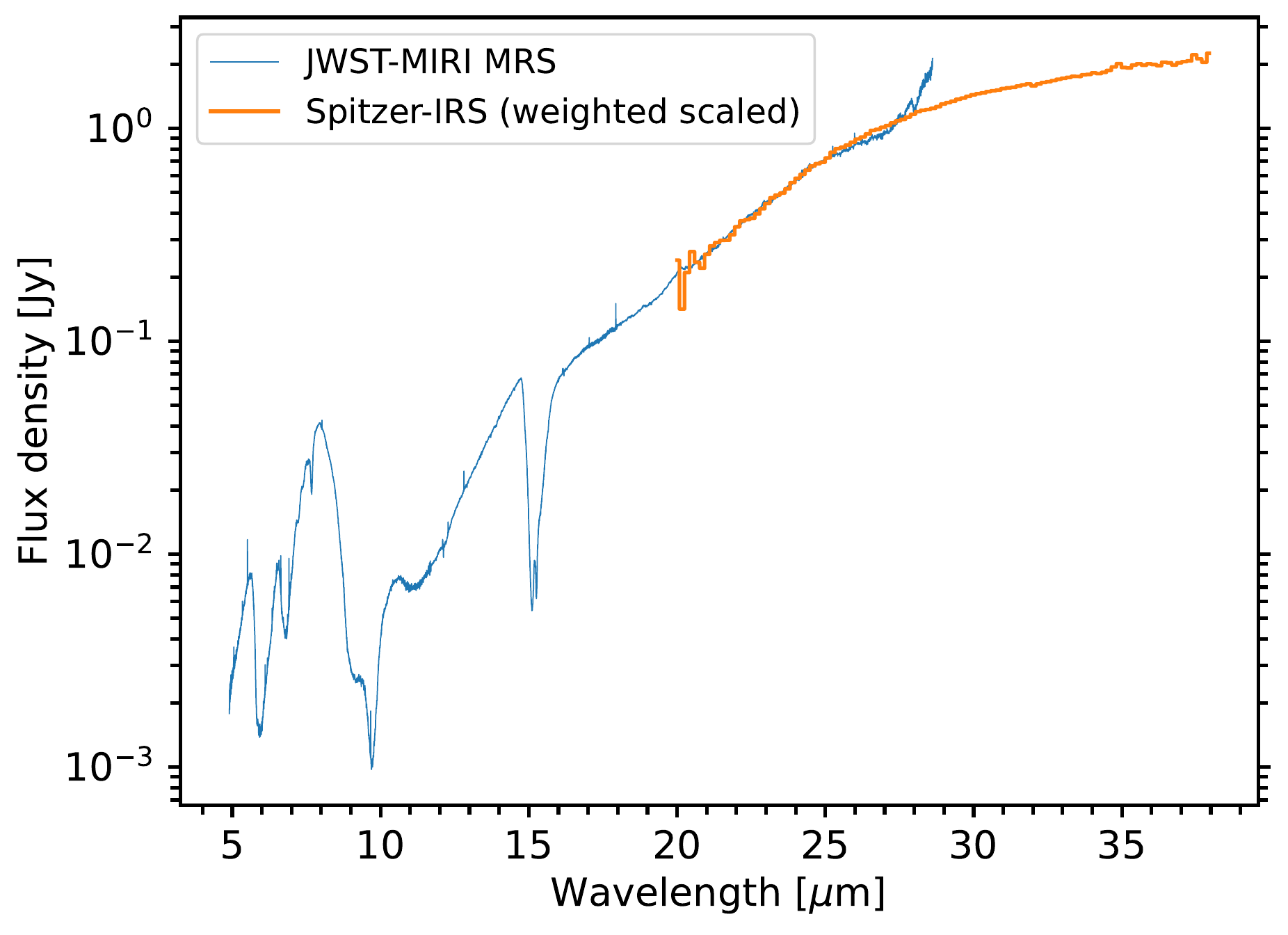}
  \caption{\textbf{Left}: The MIRI MRS spectra compared with Spitzer/IRAC aperture photometry (black). Both data were extracted with a 3\arcsec\ aperture. The red square shows the MRS spectra convolved with the IRAC 4 filter. The errorbars indicate an assumed 10\%\ uncertainty. The aperture photometry is aperture corrected but without color corrections. \textbf{Right}: The MRS 1D spectra along with the scaled Spitzer/IRS spectra at long wavelengths.  The scaling factor increases linearly with wavelength to match the wavelength-dependency in the extraction aperture of the MRS 1D spectra.}
  \label{fig:irs_comp}
\end{figure}

To check the accuracy of our overall calibration, we compared the MIRI spectra with Spitzer/IRAC aperture photometry, both extracted with a 3\arcsec\ aperture (Figure\,\ref{fig:irs_comp}). Appropriate aperture corrections were applied to the IRAC aperture photometry \citep[Table 4.8 in][]{IRAC_handbook}. After convolving the MRS spectra with the IRAC 4 filter, the spectro-photometric flux at 8 \micron\ agrees with the IRAC 4 flux. The MRS spectra have limited wavelength coverage that prevents a similar comparison at 5.8 \micron.  Figure\,\ref{fig:irs_comp} (right) shows the MRS 1D spectra extracted from the protostar compared with the scaled Spitzer/IRS Long-Low (LL1) spectra.  The IRS LL1 spectrum matches the long wavelength part of the MRS spectra, making the $\lambda > 30$ \micron\ in the IRS spectra suitable for baseline fitting.

\begin{figure*}[htbp!]
  \centering
  \includegraphics[width=\textwidth]{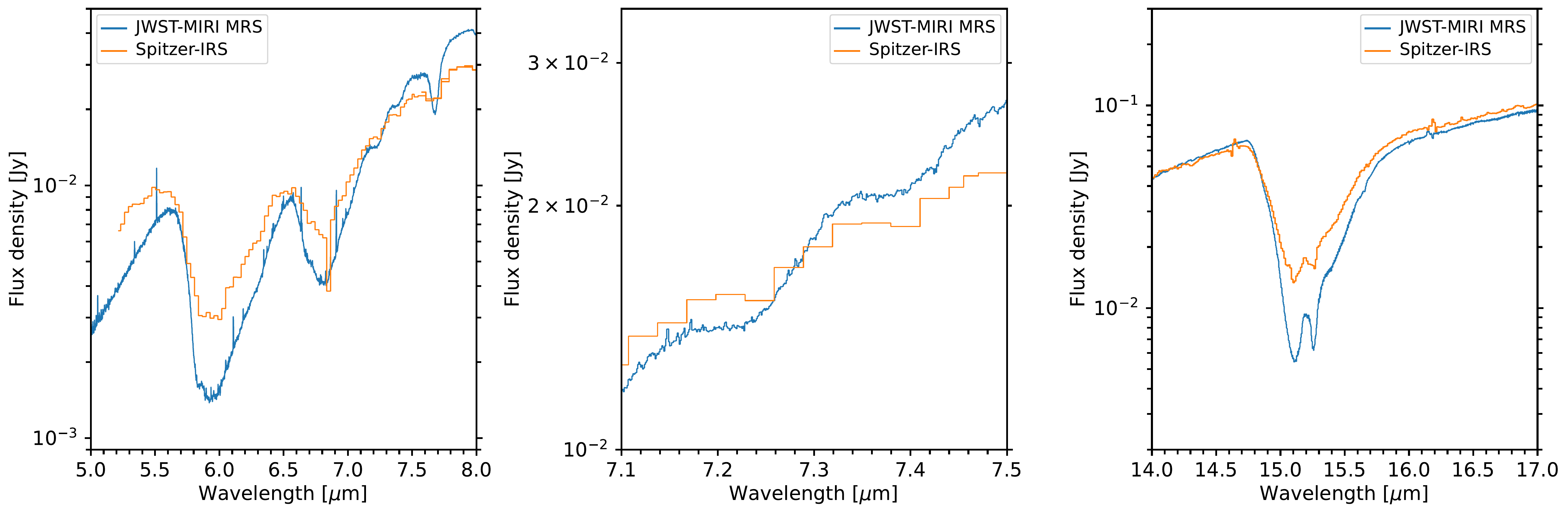}
  \caption{Selected absorption features detected in the MIRI MRS spectra compared with archival Spitzer/IRS spectra.}
  \label{fig:irs_comp_ice}
\end{figure*}

Figure\,\ref{fig:irs_comp_ice} shows the absorption features in both the MIRI MRS spectrum and the Spitzer/IRS spectra.  All features are deeper and much better resolved in the MRS spectra.  The apparent shifts in the CO$_2$ feature (15.2 \micron) may be due to the uncertainty in the wavelength solution (Section\,\ref{sec:observations}).

\section{Laboratory Data}
\label{sec:lab_data}
Several laboratory absorbance spectra are taken from the Leiden Ice Database for Astrochemistry (LIDA; \citealt{2022arXiv220812211R}) along with others that are collected from individual studies.  Table\,\ref{tbl:lab_ref} shows the references of ice species included in the composite synthetic ice spectra (Section\,\ref{sec:composite}).  Table\,\ref{tbl:com_ice_lab} lists the absorption features of organic ice species used for the discussion in Section\,\ref{sec:ice}.

\begin{deluxetable}{ccc}
    \centering
    \tablecaption{References of laboratory spectra}
    \label{tbl:lab_ref}
    \tablehead{\colhead{Species} & \colhead{Temperature (K)} & \colhead{References}}
    \startdata
      GCS 3\tablenotemark{a}                          & \nodata\ & \citet{2004ApJ...609..826K} \\
      H$_2$O                                          & 15  & \citet{2007AA...462.1187O} \\
      H$_2$O+CH$_3$OH+CO$_2$+CH$_4$ (0.6:0.7:1.0:0.1) & 10  & \citet{1999AA...350..240E} \\
      H$_2$O+HCOOH (1:1)                              & 15  & \citet{2007AA...470..749B} \\
      CH$_3$OH                                        & 15  & \citet{2018AA...611A..35T} \\
      CO$_2$                                          & 15  & \citet{2006AA...451..723V} \\
      CH$_3$CHO                                       & 15  & \citet{2018AA...611A..35T} \\
      CH$_3$CH$_2$OH                                  & 15  & \citet{2018AA...611A..35T} \\
      NH$_3$                                          & 10  & \citet{2003AA...399..169T} \\
      H$_2$CO                                         & 10  & \citet{1996AA...312..289G} \\
    \enddata
    \tablenotetext{a}{The GCS 3 spectra are taken from the ice library of ENIIGMA \citep{2021AA...654A.158R}.}
\end{deluxetable}

\begin{deluxetable}{llll}
  \centering
  \tablecaption{Complex organic ice features measured in laboratory}
  \label{tbl:com_ice_lab}
  \tablehead{\colhead{Species} & \colhead{Mode} & \colhead{Peak position} & \colhead{Reference} \\
             \colhead{}        & \colhead{} & \colhead{(\micron)} & \colhead{} }
  \startdata
  \multirow{4}{1in}{Acetaldehyde (CH$_3$CHO)}  & CH$_3$ rock. + CC stretch. + CCO bend. & 8.909 & \multirow{4}{*}{\citet{2018AA...611A..35T}} \\
                                                & CH$_3$ sym-deform. + CH wag.           & 7.427 & \\
                                                & CH$_3$ deform.                         & 6.995 & \\
                                                & C=O stretch.                           & 5.803 & \\
  \hline
  \multirow{6}{1in}{Ethanol (CH$_3$CH$_2$OH)}  & CC stretch.        & 11.36 & \multirow{6}{*}{\citet{2018AA...611A..35T}} \\
                                                & CO stretch.        & 9.514 & \\
                                                & CH$_3$ rock.       & 9.170 & \\
                                                & CH$_2$ torsion.    & 7.842 & \\
                                                & OH deform.         & 7.518 & \\
                                                & CH$_3$ sym-deform. & 7.240 & \\
  \hline
  \multirow{5}{1in}{Methyl formate (HCOOCH$_3$)} & C=O stretch.       & 5.804 & \multirow{5}{*}{\citet{2021AA...651A..95T}} \\
                                                  & C--O stretch.      & 8.256 & \\
                                                  & CH$_3$ rock.       & 8.582 & \\
                                                  & O--CH$_3$ stretch. & 10.98 & \\
                                                  & OCO deform.        & 13.02 & \\
  \enddata
  \tablecomments{The listed features are measured from amorphous ice at 15 K.}
\end{deluxetable}


\end{document}